\documentclass[
 reprint,
 amsmath,amssymb,
 aps,
 prl,
 superscriptaddress,
]{revtex4-2}

\usepackage{graphicx}
\usepackage{dcolumn}
\usepackage{bm}
\usepackage{todonotes}
\usepackage{microtype}
\usepackage{booktabs}
\usepackage{array}
\usepackage{siunitx}
\usepackage{mhchem}
\usepackage{multirow}
\usepackage[hidelinks]{hyperref}
\usepackage{bm}
\usepackage{microtype}
\usepackage{xcolor}

\usepackage{todonotes}

\makeatletter
\def\input@path{{./}{../}}
\makeatother
\graphicspath{{../}}

\newcommand{\tsr}[1]{\ensuremath{\bm{#1}}}
\renewcommand{\vec}[1]{\ensuremath{\bm{#1}}}

\renewcommand{\d}{\mathrm{d}}

\newcommand{\dd}[2]{\mathchoice{\frac{\d #1}{\d #2}}{\d #1/\d #2}{\d #1/\d #2}{\d #1/\d #2}}

\newcommand{\Tr}[1]{\ensuremath{\mathrm{Tr}\!\left(#1\right)}}

\newcommand{\cref}{c_\mathrm{ref}}

\newcommand{\beq}{\begin{equation}}
\newcommand{\eeq}{\end{equation}}
\newcommand{\beqa}{\begin{eqnarray}}
\newcommand{\eeqa}{\end{eqnarray}}
\newcommand{\beqas}{\begin{eqnarray*}}
\newcommand{\eeqas}{\end{eqnarray*}}

\DeclareMathOperator{\tr}{tr}
\makeatletter
\newcommand{\currentfsize}{\f@size pt}
\makeatother
\usepackage{xr-hyper}
\usepackage{cleveref}

\graphicspath{{}{./}}

\newcommand{\sym}{\operatorname{sym}}
\newcommand{\pos}[1]{\left(#1\right)_{+}}
\newcommand{\SC}{\mathrm{SC}}
\newcommand{\BCC}{\mathrm{BCC}}
\newcommand{\FCC}{\mathrm{FCC}}
\newcommand{\eR}{\mathbf e_r}
\newcommand{\sphereav}[1]{\overline{#1}}
\newcommand{\Rgas}{\mathcal R}

\begin{document}

\title{Battery open-circuit voltage is not purely chemical}

\author{Andrea Giudici}
\email{andrea.giudici@maths.ox.ac.uk}
\affiliation{
Mathematical Institute, University of Oxford,
Oxford OX2 6GG, United Kingdom
}

\author{Christoph Pohl}
\affiliation{
Weierstrass Institute for Applied Analysis and Stochastics (WIAS),
10117 Berlin, Germany
}

\author{Alberto Salvadori}
\affiliation{
Department of Mechanical and Industrial Engineering,
University of Brescia,
25123 Brescia, Italy
}

\author{Colin Please}
\affiliation{
Mathematical Institute, University of Oxford,
Oxford OX2 6GG, United Kingdom
}

\author{Manuel Landstorfer}
\affiliation{
Weierstrass Institute for Applied Analysis and Stochastics (WIAS),
10117 Berlin, Germany
}

\author{Jon Chapman}
\affiliation{
Mathematical Institute, University of Oxford,
Oxford OX2 6GG, United Kingdom
}

\date{\today}

\begin{abstract}
Open-circuit-voltage (OCV) curves are commonly treated as intrinsic chemical properties of electrode materials. However, this view is incomplete. In ion-insertion batteries, OCV also depends on mechanical state and microstructure. Using finite-element simulations and asymptotic analysis, we show that particle swelling and external loads promote particle--particle contact that generates compressive stresses, shifting the inserted-ion chemical potential. The OCV correction is nonlinear, follows Hertzian contact scaling, and depends on particle arrangement. Identical materials can therefore exhibit different OCV curves in different electrode microstructures. Furthermore, in full cells, electrodes are mechanically coupled through the common stack stress. Thus, cell OCV is a chemo-mechanical property of the entire battery architecture, not chemistry alone.
\end{abstract}

\maketitle


\begin{figure}[b]
    \includegraphics{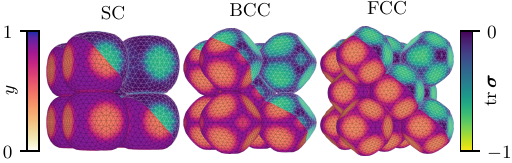}
    \caption{Simple-cubic (SC), body-centred cubic (BCC), and face-centred cubic (FCC) particle arrangements. Particle swelling upon ion-insertion introduces contact causing stress, shown by $\Tr{\bf{\sigma}}$, and lithium redistribution, shown by $y$.}
      \label{fig:1}
\end{figure}

\begin{figure*}[hbt]
  \includegraphics{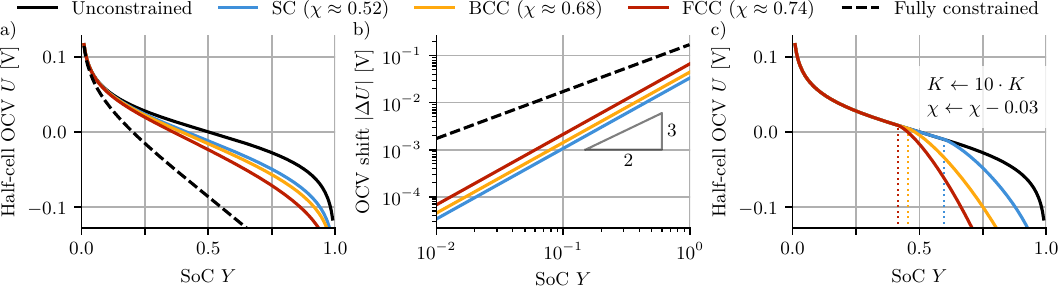}
  \caption{(a) Electrode OCV curves for unconstrained particles, fully constrained particles, and the three lattices. (b) Contact-induced OCV corrections, showing the Hertzian $\mathrm{SoC}^{3/2}$ scaling. (c) OCV curves for a stiffer material with a lowered active material volume fraction $\chi$, causing initial contact at different SoC values. For the chemical part of the chemical potential, $\mu_\mathrm{chem}=RT\ln \frac{y}{1 - y}$ was used.}
  \label{fig:numerical1}
\end{figure*}

Ion-insertion batteries store energy by transferring ions between two electrodes. At equilibrium, the voltage of a cell comprising an insertion anode $a$ and cathode $c$ is its open-circuit voltage (OCV). For monovalent ions, it is 
\begin{align}
    U_{\mathrm{cell}}
    =-\frac{1}{F}\left(\mu_c-\mu_a\right),
\end{align}
where $\mu_a$ and $\mu_c$ are the insertion chemical potentials in the anode and cathode respectively, and $F$ is Faraday's constant.
The contribution of an individual electrode is characterised by its half-cell OCV, measured against an electrode whose chemical potential is taken to be constant.
For electrode $a$, this is given by $U_a(y)=-\frac{1}{F}\mu_a(y)+\mathrm{const}$, where $y$ is the mole fraction of the inserted ion.

%
Half-cell OCV curves are key inputs in battery models~\cite{doyle1993,chen2020} and are usually treated as purely chemical properties that only depend on $y$. However, this view is incomplete. It has been shown that cell voltage, impedance, and ageing respond to
applied stack pressure~\cite{mussa2018}, while direct measurements on
insertion materials show that stress can shift OCV by
several millivolts~\cite{jiang2023effects}. These observations raise the question of whether the mechanical state of the system should be considered when determining OCV.

This question becomes particularly relevant when comparing reported OCV curves for nominally similar electrode chemistries. Such curves can differ substantially, with particularly pronounced discrepancies near stoichiometric limits ~\cite{SM}. Although material grade, ageing, hysteresis, and experimental protocol may explain part of this spread, here we show that electrode mechanics and microstructure are additional determinants of OCV and should therefore be accounted for when OCV curves are measured, compared, and used in battery models.

Electrodes are dense assemblies of active particles, through which stack pressure and particle swelling generate compressive loads that are transmitted through the electrode. These loads alter the work required to insert an ion, thereby shifting the insertion chemical potential and hence the electrode OCV. Although active-particle mechanics has been studied extensively~\cite{christensen2006,zhang2007,cheng2010,bower2011}, most existing models treat particles as mechanically isolated from the surrounding electrode. In these situations, stresses arise primarily from transient concentration gradients and relax as the concentration homogenises. Consequently, these models cannot capture stresses sustained at equilibrium or predict a change in the resulting OCV.

Continuum models have shown that externally imposed or matrix-mediated stresses can modify equilibrium voltage~\cite{bower2011,jung2025,giudici2026incorporating}. In a densely packed electrode, however, direct particle--particle contact provides a distinct load-transfer mechanism which produces larger nonlinear stresses, with strength set by the contact network. Therefore, the number of neighbouring particles, contact orientation, and packing geometry affect the electrode OCV, making microstructural organisation an important variable.

We show that particle-particle contact produces a nonlinear, microstructure-dependent OCV correction, governed, for small swelling strains, by Hertzian mechanics. The same contact law gives a nonlinear electrode stress--strain response and couples electrodes through the common stack stress. Full-cell simulations show voltage shifts of order tens of millivolts, largest near full charge.

\begin{figure}[hb]
    \includegraphics{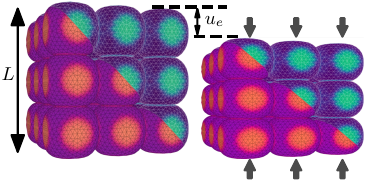}
    \caption{Imposing through-cell strain $\varepsilon_e = u_e / L$.}
    \label{fig:strain_sketch}
\end{figure}


We first consider the particle scale. For definiteness, we use a graphite electrode in a lithium-ion cell, although the mechanism applies to any insertion material undergoing small induced strains. We model the electrode as a periodic array of identical spherical particles of radius $R$ in SC, BCC, or FCC arrangements, shown in Fig.~\ref{fig:1}. The ideal lattices isolate how contact coordination and orientation affect OCV.

Let $n(\bm x,t)$ denote the number density of intercalated ions in the active material,  $n_\ell$ the number density of lattice sites, and  $y(\bm x,t) = \tfrac{n(\bm x,t)}{n_\ell}$ the mole fraction, with the average value in the particle domain $\mathcal B$,
\begin{equation}
Y=\frac{1}{|\mathcal B|}\int_{\mathcal B}y\,\mathrm dV~.
\label{eq:average_filling}
\end{equation}
Here, $Y$ can be considered as state of charge (SoC) of a single electrode. 
%
We assume isotropic linear swelling,
\begin{equation}
\boldsymbol\varepsilon_{\mathrm{sw}}=y  g\mathbf I,
\end{equation}
where $g$ is the linear swelling factor. 
The elastic strain and free-energy density are
\begin{align}
\boldsymbol\varepsilon_{\mathrm{el}}
&=\tfrac{1}{2}\left(\nabla\mathbf u+\nabla\mathbf u^T\right)
-\boldsymbol\varepsilon_{\mathrm{sw}},
\\
\!\!\psi(y,\boldsymbol\varepsilon_{\mathrm{el}})
&=\psi_{\mathrm{chem}}(y)
+\mu_L\operatorname{tr}(\tsr{\varepsilon}_\mathrm{el} \cdot \tsr{\varepsilon}_\mathrm{el})
+\tfrac12 \lambda_L \operatorname{tr}(\boldsymbol\varepsilon_{\mathrm{el}})^2.
\end{align}
where $\lambda_L$ and $\mu_L$ are the Lamé constants for the linear elastic energy. We obtain the stress tensor $\boldsymbol\sigma$ and the chemical potential $\mu$ from the free energy as 
\begin{align}
\label{eq:hook}
\boldsymbol\sigma &:= \frac{\partial \psi}{\partial \boldsymbol\varepsilon_{\mathrm{el}}}
=2\mu_L\boldsymbol\varepsilon_{\mathrm{el}}
+\lambda_L\operatorname{tr}(\boldsymbol\varepsilon_{\mathrm{el}})\mathbf I,
\\
\mu &:= \frac{\partial \psi}{\partial n} = n_\ell^{-1} \frac{\partial \psi}{\partial y}
=\mu_{\mathrm{chem}}(y)
- n_\ell^{-1} g \operatorname{tr}(\boldsymbol\sigma),
\label{eq:chemical_potential}
\end{align}
Note that the partial molar volume is
$v_c := \partial\mu/\partial p$, where
$p=-\tfrac{1}{3}\operatorname{tr}\boldsymbol{\sigma}$ is the pressure.
Equation~(7) therefore gives $g=\tfrac{1}{3}n_\ell v_c$.
Mechanical and chemical equilibrium require
\begin{equation}
\nabla\!\cdot\!\boldsymbol\sigma=0,
\qquad
\mu=\text{constant}.
\label{eq:equilibrium}
\end{equation}
For a stress-free state, we have $\boldsymbol\sigma=0$ and $\mu=\mu_\mathrm{chem}(y)$, which yields the  typical representation of the OCV. However, due to mechanical constraints upon swelling, the OCV shift 
is $\Delta U:=-\Delta\mu/F := - (\mu- \mu_\mathrm{chem}(y))/F$.   
The stress-free, empty lattice is taken as the reference state at
$Y=Y_0=0$, corresponding to zero mole fraction. 
In this state, the neighbouring particles are assumed to be just touching at $Y=0$ (see SI ~\cite{SM} for details on alternative initial conditions). We assume further that particle surfaces are traction-free and interact through frictionless contact.

A fully constrained (fc) particle (i.e. vanishing displacement) provides a lower bound to the OCV.
In this case, at equilibrium, $y=Y$ and $\boldsymbol\varepsilon_{\mathrm{el}}=- y g \mathbf I$, yielding
\begin{equation}
\Delta U_{\mathrm{fc}} = - 9 \tfrac{K}{F n_\ell}  g^2 Y ~, 
\qquad
K =\lambda_L+\tfrac{2}{3}\mu_L.
\label{eq:fully_constrained}
\end{equation}
In an electrode, confinement is instead localised at particle
contacts. We solve Eqs.~\eqref{eq:hook}--\eqref{eq:equilibrium}
numerically in the three particle arrangements (see Fig.~\ref{fig:1}), 
accounting for contact upon swelling by a penalty method ~\cite{SM}, 
and with no relative motion of particles, equivalent to a strain-free clamped electrode. Fig.~\ref{fig:numerical1} shows that contact produces residual stress and lithium redistribution even at equilibrium. The resulting OCV correction due to swelling is several millivolts and scales as $\mathrm{SoC}^{3/2}$.

\begin{figure*}[t]
  \includegraphics{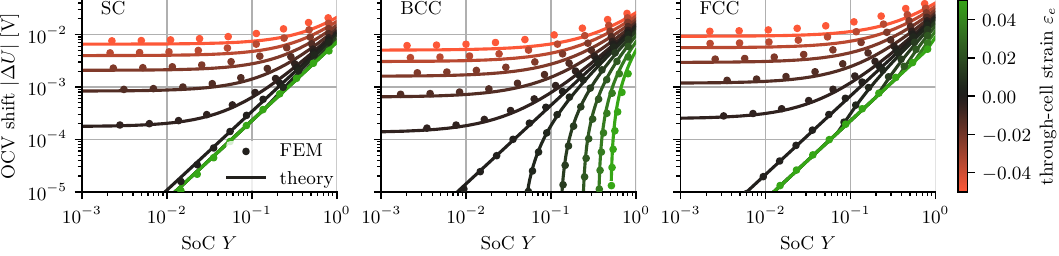}
  \caption{Contact-induced OCV correction from finite-element simulations (dots) and Eq.~\eqref{eq:dmu_lattice} (lines) for SC, BCC, and FCC lattices over a range of imposed through-cell strains $\varepsilon_e$.}
  \label{fig:theory_ocv_comparison}
\end{figure*}

Electrode-scale deformation can also change the correction.
Since current collectors strongly restrict in-plane deformation, we only consider through-cell strain $\varepsilon_e$~\cite{giudici2025mechanical} by allowing particles centres to move in the through-cell direction, see Fig. ~\ref{fig:strain_sketch}.
Fig.~\ref{fig:theory_ocv_comparison} shows that at a given $Y$, changing
$\varepsilon_e$ changes the contact state, hydrostatic stress, and
OCV. The simulations show that chemical composition  is not sufficient to determine OCV: swelling, electrode
strain, and microstructure composition all enter the problem.

The numerical results can be understood through a weak-contact
expansion. Let $\mathbf e$ denote the through-cell direction and
$\mathbf n_m$ the unit normal vector of the $m$-th contact plane.
We take $g\ll1$ and
$\varepsilon_e=O(g)$. At leading order the particle undergoes
homogeneous stress-free swelling, so $y=Y$. The indentation of contact
$m$ is therefore
\begin{equation}
\delta_m
=
2R\left(gY-\alpha_m\varepsilon_e\right)_+,
\qquad
\alpha_m=(\mathbf n_m\cdot\mathbf e)^2,
\label{eq:indentation}
\end{equation}
where $(x)_+=\max(x,0)$ and $\alpha_m$ is the squared projection of the contact normal onto the through-cell direction. Classic Hertz contact theory~\cite{johnson1987contact} gives
\begin{align}
\label{eq:Pn}
P_m
=
\tfrac{4}{3}E^*\sqrt{\tfrac12 R}\,\delta_m^{3/2}, \quad
E^*
=
\frac{2\mu_L(\lambda_L+\mu_L)}
     {\lambda_L+2\mu_L}.
\end{align}
Thus $P_m=O(g^{3/2})$: free swelling enters at $O(g)$, whereas
the first contact load enters at $O(g^{3/2})$.

Even with knowledge of the contact loads, finding the particle
deformation and $y$ distribution remains a fully 3D problem that
cannot be solved analytically. However, we notice that the radial
symmetry is broken only by the contact load, which scales like
$g^{3/2}$. Therefore, we expect $y=\bar{y}+g^{3/2} \delta y$, where $\delta y$ is $O(1)$ and
overbar denotes spherical mean, defined in spherical coordinates $r,\theta,\phi$, as 
\begin{equation}
\overline{\eta}(r)
=
\frac{1}{4\pi}
\int_{\mathbb S^2}\eta(r,\theta,\phi)\,\sin\theta \,\mathrm{d}\theta \, \mathrm{d}\phi.
\end{equation} 

We now expand $\mu_\mathrm{chem}$ in small $g$, use the constitutive relation in \eqref{eq:hook} into both
Eqs.~\eqref{eq:chemical_potential} and \eqref{eq:equilibrium}, and
average over the surface to obtain equilibrium conditions 
\begin{equation}
\mu_{\mathrm{chem}}(\overline{y})
-\frac{v_c}{3}(2\mu_L+3\lambda_L)
\left(
\overline u'
+\frac{2}{r}\overline u
-3g\overline y
\right)
=\mu,
\label{eq:average_mu}
\end{equation}
\begin{equation}
\overline u''
+\frac{2}{r}\overline u'
-\frac{2}{r^2}\overline u
=
\frac{2\mu_L+3\lambda_L}{2\mu_L+\lambda_L}
\,g\,\overline y',
\label{eq:average_mech}
\end{equation}
where prime identifies derivative with respect to $r$ and we used that $\overline{\delta y}=0$. 
The averaged normal traction condition at the particle surface is given by
\begin{equation}
(2\mu_L+\lambda_L)\overline u'
+\frac{2\lambda_L}{R}\overline u
-(2\mu_L+3\lambda_L)g\overline y
=
-\frac{1}{4\pi R^2}\sum_mP_m,
\label{eq:average_bc}
\end{equation}
where the sum over contact points $P_m$ is given in Eq.~\eqref{eq:Pn}. We also require the displacement at the centre to vanish. Finally, the constraint in Eq.~\eqref{eq:average_filling} becomes
\begin{equation}
\frac{3}{R^3}\int_0^R\overline y\,r^2\,\mathrm dr=Y.
\label{eq:average_filling_radial}
\end{equation}

The problem \eqref{eq:average_mu}-\eqref{eq:average_filling_radial} has solution
\begin{equation}
\overline y=Y,
\quad
\overline u(r)
=
\left(
gY-\frac{\frac{1}{4\pi R^2}\sum_mP_m}{2\mu_L+3\lambda_L}
\right)r,
\label{eq:averaged_solution}
\end{equation}
\begin{equation}
\label{eq:mu_contact_sum}
\mu=\mu_{\mathrm{chem}}(Y)+\frac{v_c}{4\pi R^2}\sum_m P_m.
\end{equation}
A full rigorous derivation is available in the SI ~\cite{SM}. Our results show that ion redistribution affects the local fields but not the leading particle-averaged chemical-potential shift, which depends only on the
closed contact loads.


For SC, four contacts have $\alpha_m=0$ and two have $\alpha_m=1$.
For BCC, all eight contacts have $\alpha_m=\tfrac13$. For FCC, four contacts 
have $\alpha_m=0$ and eight have $\alpha_m=\tfrac12$. Substitution into
Eq.~\eqref{eq:mu_contact_sum} gives the leading-order OCV correction
\begin{equation} 
\Delta U^{\mathcal L} =
-\frac{2E^*v_c}{3\pi F}
\left[
a_{\mathcal L}(gY)^{3/2}
+
b_{\mathcal L}
\left(gY-c_{\mathcal L}\varepsilon_e\right)_+^{3/2}
\right],
\label{eq:dmu_lattice}
\end{equation}
where $(a_{\mathcal L},b_{\mathcal L},c_{\mathcal L})$ is
$(4,2,1)$, $(0,8, \tfrac13)$, or $(4,8, \tfrac12)$ for
$\mathcal L=\mathrm{SC}$, BCC, or FCC. Fig.~\ref{fig:theory_ocv_comparison} shows our result is in excellent agreement with the
finite-element calculations.

\begin{figure*}[ht]
  \includegraphics{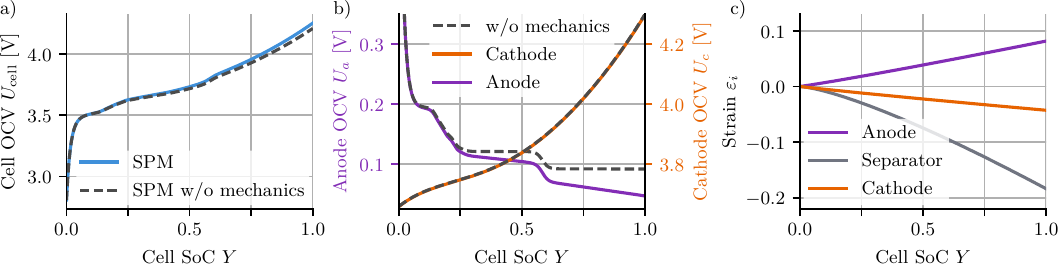}
  \caption{PyBaMM simulation of a clamped cell, $\varepsilon_{\mathrm{cell}}=0$, using a SPM model. Comparison of full-cell (a) and electrode OCV (b) with and without the contact correction.  (c) Strain evolution in the different cell elements.}
  \label{fig:pybammSim}
\end{figure*}

Equation (19) is the general result. It shows that contact introduces an OCV correction which depends nonlinearly on swelling and electrode strain, and explicitly on microstructural geometry. Thus, even at fixed chemistry and ion content, electrodes with different particle arrangements can have different OCVs. In a disordered or polydisperse electrode, the lattice constants are replaced by the distributions of contact number, orientation, indentation, and force. Binder and conductive additives can carry part of the load \cite{giudici2026incorporating}, but they do not remove the dominant contact contribution. Electrode processing is known to change coordination and contact anisotropy~\cite{stershic2015,sangros2019}; Eq.~\eqref{eq:mu_contact_sum} shows that these same structural descriptors also enter equilibrium OCV.
This provides a mechanical interpretation for part of the spread among
measured electrode OCV curves. Curves reported for nominally similar
graphite and layered-oxide electrodes differ most strongly near the
ends of their usable stoichiometric
ranges~\cite{ecker2015,chen2020,li2021ocv}. Those differences need not
be mechanical, and ageing or changes in active-material composition
can themselves reshape OCV~\cite{schmitt2021}. However, our result shows that microstructure and mechanical constraints can also alter OCV curves.

To understand how particle contact affects the electrode scale mechanics, we derive its stress-strain relationship. Averaging the contact forces crossing a plane normal to the through-cell direction we find
\begin{equation}
\sigma_e^{\mathcal L}
=-\frac{2E^*}{3}A_{\mathcal L}
\left(B_{\mathcal L}gY-C_{\mathcal L}\varepsilon_e\right)_+^{3/2},
\label{eq:stress_lattice}
\end{equation}
where $(A_{\mathcal L},B_{\mathcal L},C_{\mathcal L})$ is $(1,1,1)$, $(\sqrt{3},1, \tfrac13)$, or $(1,2,1)$ for SC, BCC, or FCC. This nonlinear stress--strain law transmits particle-scale contact mechanics to the cell stack.

Consider a positive electrode, separator, and negative electrode with thicknesses $h_c$, $h_s$, and $h_a$. We require
\begin{equation}
h_c\varepsilon_c+h_s\varepsilon_s+h_a\varepsilon_a
=H\varepsilon_{\mathrm{cell}},
\qquad
H=h_c+h_s+h_a,
\label{eq:stack_compatibility}
\end{equation}
while mechanical equilibrium imposes common stress \cite{giudici2025mechanical},
\begin{equation}
\sigma=\sigma_c=\sigma_s=\sigma_a.
\label{eq:stack_equilibrium}
\end{equation}
We take the separator to be linear elastic, $\sigma_s=M_s\varepsilon_s$, and use Eq.~\eqref{eq:stress_lattice} for each electrode. Eqs. ~\eqref{eq:stack_compatibility}--\eqref{eq:stack_equilibrium} then determine the electrode strains and contact loads for prescribed electrode mole fractions and cell strain. A pressure-controlled cell follows from prescribing $\sigma$ instead.
The voltage correction of electrode $k$ is
\begin{equation}
\Delta U_k=-\frac{1}{zF}
\frac{v_{c,k}}{4\pi R_k^2}
\sum_{n\in k}P_{n,k},
\end{equation}
and the full-cell correction is
\begin{equation}
\Delta U_{\mathrm{cell}}
=\Delta U_c-\Delta U_a
=-\frac{\Delta\mu_c-\Delta\mu_a}{zF}.
\label{eq:cell_voltage_shift}
\end{equation}
Thus swelling in either electrode changes the common stress and can shift the OCV contribution of both.

We implement this coupled model in PyBaMM~\cite{sulzer2021}. Fig.~\ref{fig:pybammSim} shows results for a clamped cell, $\varepsilon_{\mathrm{cell}}=0$, comparing the mechanical correction to the OCV with standard SPM results for an FCC lattice. The full-cell OCV increases by tens of millivolts near full charge.  The change is driven primarily by mechanics in the negative electrode due to the larger partial molar volume of lithium in graphite. 

We have shown that identical active materials can exhibit different OCV curves due to their particle arrangement and mechanical state, offering an explanation for OCV differences observed in experiments. Our simplified lattice arrangements are instructive to show how microstructure affects OCV, but more careful study is required to characterise realistic electrode composition and predict its properties.
At the full-cell scale, the common stack stress mechanically couples the two electrodes. Charging redistributes strain between the electrodes and separator, generates nonlinear stack stress, and shifts the full-cell OCV. Our analytical results, in agreement with numerics, offer a way to estimate this OCV correction. 

In conclusion, we have shown that electrode OCV curves should not be understood independently of mechanics. External loadings alter the OCV curve. More fundamentally, even in the absence of external loading, equilibrium OCV response depends on electrode microstructure due to internal stresses and is therefore not purely chemical.

\begin{acknowledgments}
A.G. and J.C. acknowledge support from the Faraday Institution
(Grant No.~FIRG095).
\end{acknowledgments}

\bibliographystyle{apsrev4-2}
\bibliography{Main_PRL_v2}

\clearpage
\onecolumngrid

\setcounter{page}{1}
\setcounter{equation}{0}
\setcounter{figure}{0}
\setcounter{table}{0}
\setcounter{section}{0}
\setcounter{subsection}{0}
\setcounter{subsubsection}{0}
\setcounter{footnote}{0}

\renewcommand{\theequation}{S\arabic{equation}}
\renewcommand{\thefigure}{S\arabic{figure}}
\renewcommand{\thetable}{S\arabic{table}}
\setcounter{secnumdepth}{3}
\renewcommand{\thesection}{S\arabic{section}}
\renewcommand{\thesubsection}{\thesection.\arabic{subsection}}
\renewcommand{\thesubsubsection}{\thesubsection.\arabic{subsubsection}}

\title{Supplemental Material for\\
``Battery open-circuit voltage is not purely chemical''}

\author{Andrea Giudici}
\email{andrea.giudici@maths.ox.ac.uk}
\affiliation{Mathematical Institute, University of Oxford, Oxford OX2 6GG, United Kingdom}

\author{Christoph Pohl}
\affiliation{Weierstrass Institute for Applied Analysis and Stochastics (WIAS), 10117 Berlin, Germany}

\author{Alberto Salvadori}
\affiliation{Department of Mechanical and Industrial Engineering, University of Brescia, 25123 Brescia, Italy}

\author{Colin Please}
\affiliation{Mathematical Institute, University of Oxford, Oxford OX2 6GG, United Kingdom}

\author{Manuel Landstorfer}
\affiliation{Weierstrass Institute for Applied Analysis and Stochastics (WIAS), 10117 Berlin, Germany}

\author{Jon Chapman}
\affiliation{Mathematical Institute, University of Oxford, Oxford OX2 6GG, United Kingdom}

\date{\today}
\maketitle
\onecolumngrid
\section{Comparison of different OCV curves reported in literature}

Open Circuit Voltage (OCV) is usually measured using galvanostatic intermittent titration technique (GITT) ~\cite{chen2020,Flores2026} or by reading near equilibrium pseudo-OCV when cycling at low C-rates (C/20 or C/50).  These measurements are important for battery modelling, simulations and  for predicting battery state of charge. Different electrode materials have different OCV curves, associated with their different chemistry. Ideally, electrodes made by the same material have the same OCV curve. However, we show that this is not always true: the same electrode material may display significantly different OCV curves in different experiments. Although these variations may be attributed to differences in experimental protocol, their root origin is still not well understood. Here, we propose that boundary conditions and microstructural composition may account at least for an important part of the discrepancy. 

Deducing the actual OCV of an intercalation material with respect to its thermodynamic mole fraction $y$ is a difficult task, since several aspects impact the measurement of the cell voltage, namely (i) the measurement protocol, GITT or pseudo-OCV during slow charging, (ii) the branch, i.e., lithiation, delithiation, or an unstated mean of the two, (iii) the voltage window together with the
active-material mass, which set the capacity by which the abscissa is normalised, and
(iv) degradation effects occurring in the first cycles. Furthermore, the mechanical conditions under which the electrochemical measurements are carried out and the specific micro-structure of an electrode material are even more rarely stated.  

We consider SINTEF OCV datasets~\cite{Flores2026} of seven graphite, six NMC111, and eight NMC532 half-cells, comprising GITT, GITT-with-hold, pseudo-OCV and pseudo-OCV-with-hold measurements. To our knowledge this is the only openly available GITT half-cell record that combines replicate
cells of a single electrode batch, several measurement protocols, and an individually
reported active mass. Our aim is to compare the OCV curves for the same material among different experimental protocols to capture the underlying differences that may arise. In particular, we take great care to address points (iii) and (iv) to make a significant comparison by correctly finding the capacity and accounting for degradation effects occurring in the initial cycles. 

To compare OCV from a thermodynamic point of view we relate the OCV to the mole fraction $y$ of
intercalated lithium. The latter is related to the capacity $Q$ via
\begin{align}
  Q = F\int_{\Omega_A} n_{A_\ell}\,y \,\d V = q_A V_A\,Y ,
  \qquad
  Y = \frac{1}{V_A}\int_{\Omega_A} y \,\d V ,
\end{align}
where $n_{A_\ell}$ denotes the molar concentration of Li lattice sites, $F$ is
Faraday's constant, $q_A = F n_{A_\ell}$ is the charge density of the material, and
$V_A = |\Omega_A|$ is the volume of the active material. Since the host deforms upon lithiation, $\Omega_A$, $n_{A_\ell}$ and $\rho_A$ are
understood in the reference configuration, for which conservation of lattice sites renders
the factorisation exact. With $V_A = M_A\rho_A^{-1}$ and $q_A^\rho = q_A\rho_A^{-1}$
[\si{\milli\ampere\hour\per\gram}] we obtain
\begin{align}
  Y = \frac{Q}{q_A^\rho M_A} ,
  \qquad
  q^\rho_\text{graphite} = \SI{371.90}{\milli\ampere\hour\per\gram},
  \quad
  q^\rho_\text{NMC111} = \SI{277.85}{\milli\ampere\hour\per\gram},
  \quad
  q^\rho_\text{NMC532} = \SI{277.58}{\milli\ampere\hour\per\gram}.
\end{align}
Note that $y \equiv Y$ holds if and only if the material is free of concentration
gradients. Coulometry thus yields only the volume average $Y$, whereas the measured
potential is governed by the occupancy at the particle surface,
$y|_{\partial\Omega_A}$. These coincide only at equilibrium, which is one reason why GITT
and pseudo-OCV measurements can diverge.

Experimentally accessible are the active mass $\hat M_A$, or equivalently the \emph{theoretical}
capacity $\hat Q_A := q_A^\rho \hat M_A$ of the specific electrode, and the charge
$\hat Q(t) = \int_0^t \hat I\,\d t'$ stored at time $t$. If the entire current flows into the intercalation reaction, i.e. if no
parasitic side reactions occur, we have $\d Q/\d t = \hat I$ and thus
\begin{align}
  \hat Y(t) = Y_0 + \frac{1}{\hat Q_A}\int_0^t \hat I\,\d t' ,
  \qquad
  Y_0 = \begin{cases} 0 & \text{graphite (\ce{C6})},\\ 1 & \text{NMC (\ce{LiMO2})},\end{cases}
  \label{eq:SI_exact}
\end{align}
where $Y_0$ is fixed by the pristine state. Equation~\eqref{eq:SI_exact} does not hold in practice, because (electronic) charge is consumed by parasitic reactions such as SEI growth or electrolyte oxidation, and hence the measured current is not just changing $\hat Y$. In general the measured current splits as
\begin{align}
  \hat I = \dd{Q}{t} + \hat I_\text{par} ,
  \label{eq:SI_current_split}
\end{align}
where $\hat I_\text{par}$ collects all non-intercalation faradaic
processes~\cite{pohlModelSEIGrowth2026,pinson2013}, so that
$\hat Y(t)-Y(t)=\hat Q_A^{-1}\int_0^t\hat I_\text{par}\,\d t'$.

\begin{figure}[t]
\centering
\includegraphics[width=\textwidth]{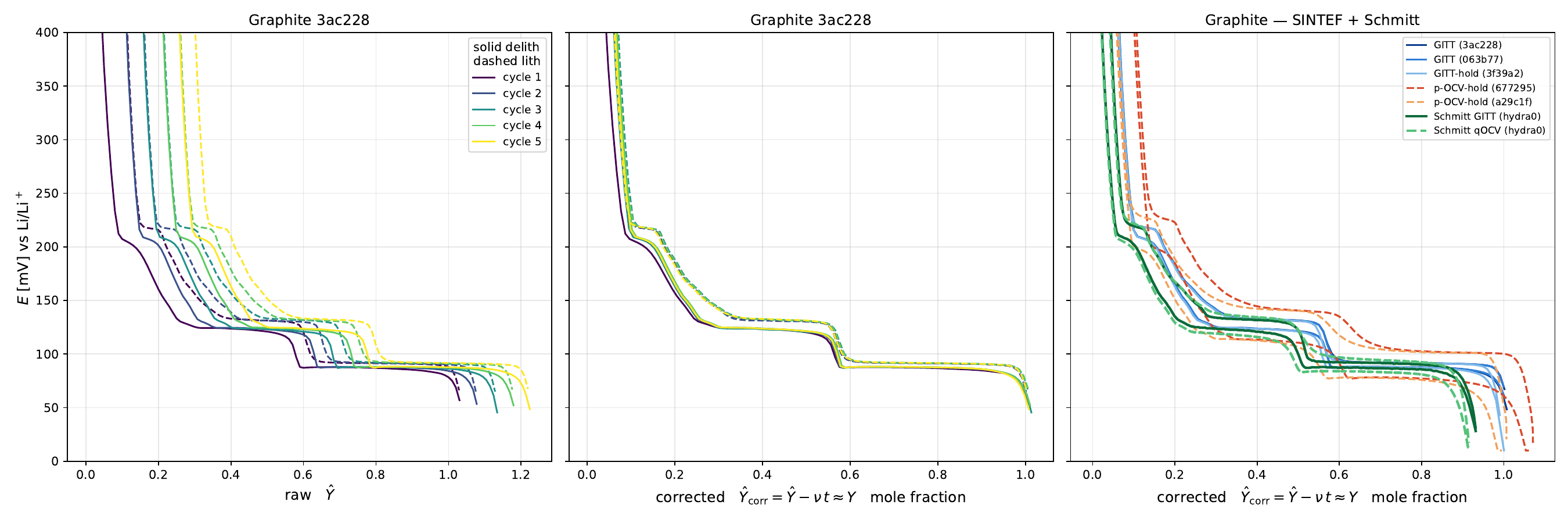}
\caption{Graphite. Left and middle: cell \texttt{3ac228}, all five cycles, before and after
the correction of Eq.~\eqref{eq:SI_corr}, solid delithiation and dashed lithiation. Right: (last) charge-discharge cycle of the five SINTEF cells (GITT, pseudo-OCV), corrected according to Eq.~\eqref{eq:SI_corr} and table \ref{tab:vertex}, showing the variation of the OCV among the cells. 
Additionally, GITT and quasi-OCV curves of Schmitt \emph{et
al.}~\cite{Schmitt2026} (uncorrected).}
\label{fig:graphite-combined}
\end{figure}

If we use \eqref{eq:SI_exact} in an uncorrected form, then for graphite this predicts $\hat Y = 1.23$ after five GITT cycles, which is impossible for \ce{LiC6} (see Figure \ref{fig:graphite-combined}, left). Fully accounting for this requires understanding the parasitic reactions, however here we can account for much of this behaviour by assuming the parasitic reaction over these few cycles occurs at a constant rate. We calculate this drift using the turning points $t_k$, i.e., the instants of current reversal terminating the first half-cycle of each cycle. At these points, both branches share the same electrode potential 
at the same instant and thus the same $\hat Y$, which the measurement confirms. Hence we introduce a correction to $\hat Y(t)$, which accounts for the parasitic reaction,  i.e.
\begin{align}
  \hat Y_\text{corr}(t) = \hat Y(t) - \nu\,t 
  \label{eq:SI_corr}
\end{align}
where $\nu$  is a constant per cell. We assume thus 
$\hat I_\text{par} = \nu\,\hat Q_A = \text{const.}$, the classical calendar-life
linearisation~\cite{broussely2001}, to be valid here over a few cycles. Note that over longer times the parasitic current is not constant and may decay as
$t^{-1/2}$~\cite{pinson2013,pohlModelSEIGrowth2026}. The quantity $\hat I_\text{par}$
is essentially the charge-endpoint slippage of high-precision
coulometry~\cite{smith2010precision,smith2010hp,bond2013}, used likewise to align
half-cell curves in degradation diagnostics~\cite{dubarry2012}. The rate $\nu$ follows
from a single linear regression on the turning points of each cell, whereby the
formation vertex is excluded. This presumes a fixed mole fraction at the cut-off
voltage, which is well satisfied for GITT and less so for pseudo-OCV. We thus consider $\hat Y_\text{corr} \approx Y$, i.e., the volume-averaged mole fraction of intercalated lithium in the active material.\\

\begin{figure}[t]
\centering
\includegraphics[width=\textwidth]{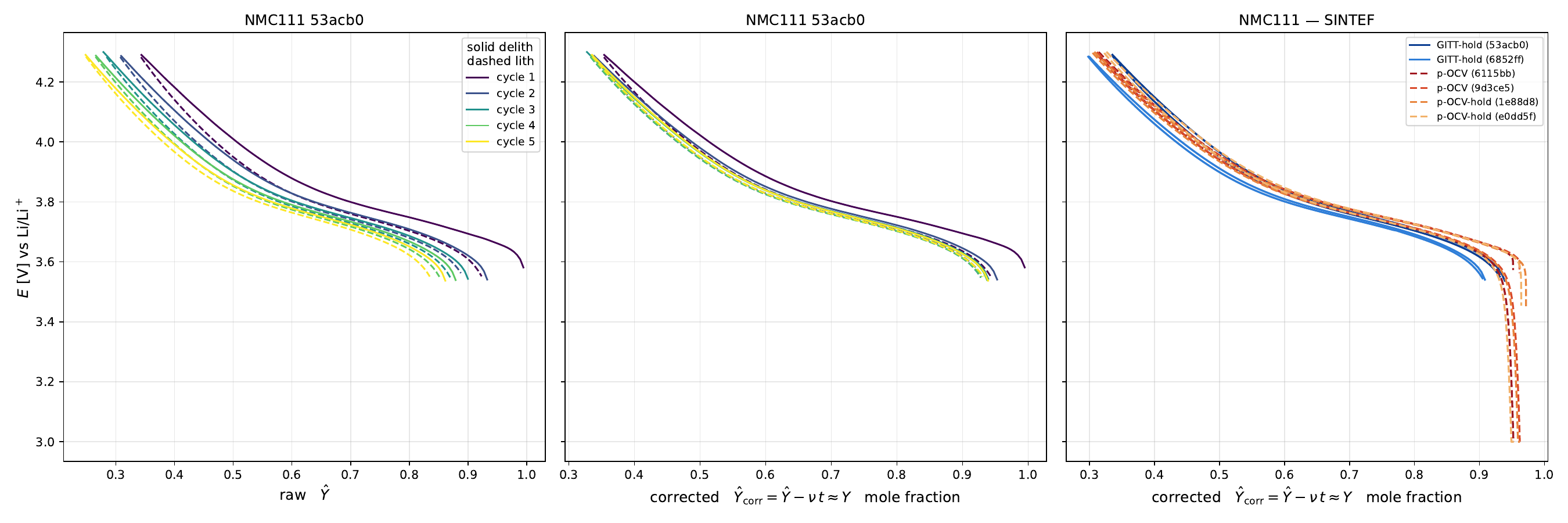}
\caption{NMC111, as Fig.~\ref{fig:graphite-combined}, with demonstration cell
\texttt{53acb0} (Left and middle). Right: Experimental NMC111 OCV of six SINTEF cells.}
\label{fig:nmc111-combined}
\end{figure}

\begin{figure}[t]
\centering
\includegraphics[width=\textwidth]{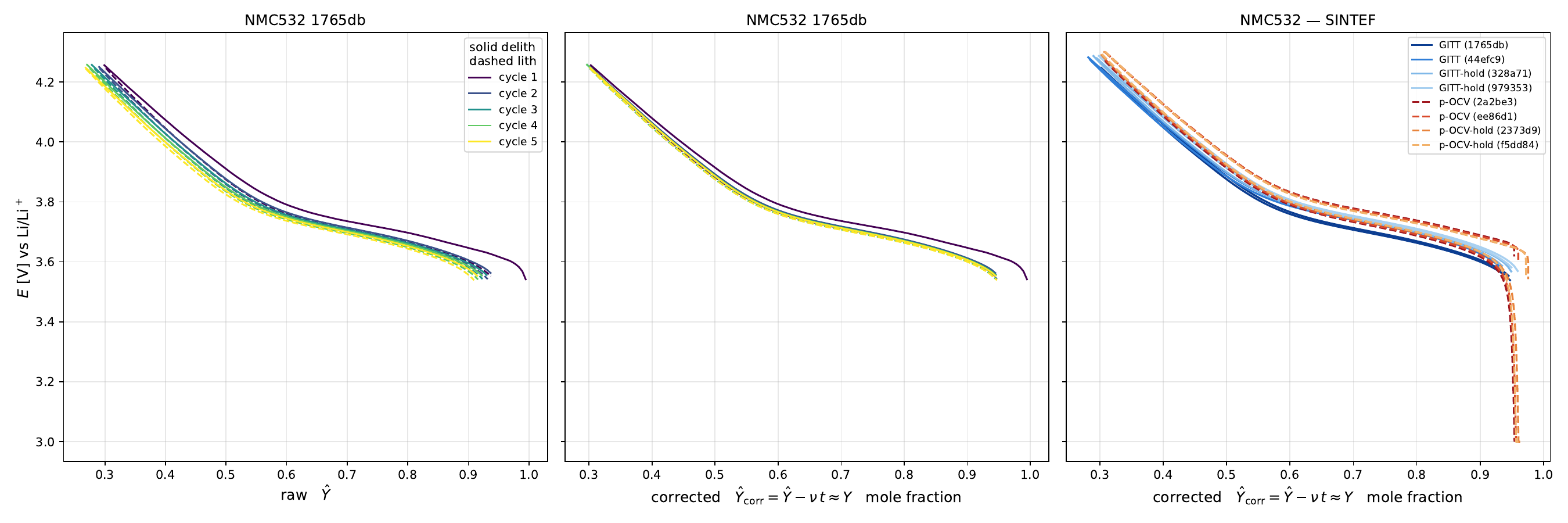}
\caption{NMC532, as Fig.~\ref{fig:nmc111-combined}, with demonstration cell \texttt{1765db} (Left and middle). Right: Experimental NMC532 OCV of eight SINTEF cells.}
\label{fig:nmc532-combined}
\end{figure}

The SINTEF data suggests $\hat I_\text{par} = +\SI{0.16+-0.03}{\micro\ampere}$
for five of the seven graphite cells and $\hat I_\text{par} = \SI{-0.10+-0.02}{\micro\ampere}$
for the four GITT-type NMC532 cells, that is $0.3$ to $0.4\,\%$ of
the pulse current, of opposite sign as the chemistry requires.
The remaining cells give larger magnitudes, see table~\ref{tab:vertex}. The graphite cycles spanning $\hat Y = 0.01$ to $1.23$ collapse onto two curves (see Figure \ref{fig:graphite-combined}), and the cells agree to a median
of \SI{0.3}{\milli\volt}, which is the relaxation floor of the measurement itself. This is a remarkable agreement, given that no parameter is fitted across cells of the same type. The same procedure is also applied to NMC111 (Figure \ref{fig:nmc111-combined}) and NMC532 (Figure \ref{fig:nmc532-combined}), showing similar \emph{convergence} across cells of the same type. This method is applied to every single cell of the SINTEF data, where table \ref{tab:vertex} summarizes the deduced values for $\hat Q_A$ and $\nu$, yielding, to the best of our knowledge, representations of the OCV w.r.t. $Y \approx \hat Y - \nu t $. Figure \ref{fig:graphite-combined} (right) then shows the OCV for graphite, Figure \ref{fig:nmc111-combined} (right) for NMC111, and Figure \ref{fig:nmc532-combined} (right) for NMC532, with variations among the different cells. \\

Two further graphite curves, GITT and
quasi-OCV, are taken from the graphite/LNMO dataset of Schmitt \emph{et
al.}~\cite{Schmitt2026}, recorded in an ECC-PAT-Core cell (EL-CELL,
\SI{18}{\milli\metre})~\cite{ElCellPATCore}. There the active mass is not reported and
$\hat Q_A$ was reconstructed from coating thickness, porosity and composition, however, without a correction $\nu$.\\

\begin{table}[t]
\centering
\caption{Derived data from SINTEF
dataset~\cite{Flores2026}. The rate $\nu$ is the slope of one linear regression on the turning points of each cell, the formation vertex being excluded, and is the only parameter
of the correction, Eq.~\eqref{eq:SI_corr}. Protocols: G = GITT, Gh = GITT-with-hold, P = pseudo-OCV, Ph = pseudo-OCV-with-hold. }
\label{tab:vertex}
\setlength{\tabcolsep}{3pt}
\resizebox{\textwidth}{!}{%
\begin{tabular}{lrrrrrrrrrrrrrrrrrrrrr}
\toprule
 & \multicolumn{7}{c}{graphite} & \multicolumn{6}{c}{NMC111} & \multicolumn{8}{c}{NMC532} \\
\cmidrule(lr){2-8}\cmidrule(lr){9-14}\cmidrule(lr){15-22}
cell & \rotatebox{90}{\texttt{3ac228}} & \rotatebox{90}{\texttt{063b77}} & \rotatebox{90}{\texttt{3f39a2}} & \rotatebox{90}{\texttt{677295}} & \rotatebox{90}{\texttt{a29c1f}} & \rotatebox{90}{\texttt{1d5628}} & \rotatebox{90}{\texttt{4ccc47}} & \rotatebox{90}{\texttt{53acb0}} & \rotatebox{90}{\texttt{6852ff}} & \rotatebox{90}{\texttt{6115bb}} & \rotatebox{90}{\texttt{9d3ce5}} & \rotatebox{90}{\texttt{1e88d8}} & \rotatebox{90}{\texttt{e0dd5f}} & \rotatebox{90}{\texttt{1765db}} & \rotatebox{90}{\texttt{44efc9}} & \rotatebox{90}{\texttt{328a71}} & \rotatebox{90}{\texttt{979353}} & \rotatebox{90}{\texttt{2a2be3}} & \rotatebox{90}{\texttt{ee86d1}} & \rotatebox{90}{\texttt{2373d9}} & \rotatebox{90}{\texttt{f5dd84}} \\
\midrule
protocol & G & G & Gh & Ph & Ph & P & P & Gh & Gh & P & P & Ph & Ph & G & G & Gh & Gh & P & P & Ph & Ph \\
$\hat Q_A$ [\si{\milli\ampere\hour}] & 2.2239 & 2.2070 & 2.1597 & 2.0514 & 2.2781 & 2.2645 & 2.1630 & 5.6765 & 5.6765 & 5.6765 & 5.6765 & 5.6765 & 5.6765 & 8.7167 & 8.7167 & 8.7167 & 8.7167 & 8.7167 & 8.7167 & 8.7167 & 8.7167 \\
$\nu$ [\SI{e-5}{\per\hour}] & $+7.999$ & $+6.463$ & $+6.201$ & $+6.776$ & $+8.493$ & $+18.68$ & $+20.11$ & $-2.791$ & $-2.982$ & $-5.275$ & $-4.985$ & $-7.086$ & $-8.510$ & $-1.026$ & $-1.423$ & $-1.034$ & $-1.202$ & $-2.138$ & $-2.891$ & $-3.849$ & $-2.494$ \\
$\hat I_\text{par}$ [\si{\micro\ampere}] & $+0.178$ & $+0.143$ & $+0.134$ & $+0.139$ & $+0.193$ & $+0.423$ & $+0.435$ & $-0.158$ & $-0.169$ & $-0.299$ & $-0.283$ & $-0.402$ & $-0.483$ & $-0.089$ & $-0.124$ & $-0.090$ & $-0.105$ & $-0.186$ & $-0.252$ & $-0.336$ & $-0.217$ \\
\bottomrule
\end{tabular}}
\end{table}

The right figures in \ref{fig:graphite-combined}, \ref{fig:nmc111-combined} and \ref{fig:nmc532-combined} show that even after a careful assessment of the OCV data of a single material, significant differences in the OCV curves among different cells can be observed. This observation suggests chemistry alone may not be sufficient to characterise OCV. As proposed in the main text, other parameters, such as mechanical boundary conditions or microstructural composition of the electrode may play a significant role in determining these curves. More systematic experimental investigations of the same material under (a) different micro-structures and (b) different macro-scale mechanical conditions can help to better understand that the battery open-circuit voltage is not purely a chemical effect.

\section{Conventions, reference state, and particle geometry}
\label{sec:SI_conventions}

Tensile stress and extensional strain are taken to be positive, whereas compression is
negative. The imposed electrode strain $\varepsilon_e$ is therefore positive
when the lattice is extended in the through-cell direction. Contact loads
$P_m$ and contact pressures $p_m$ are positive magnitudes in compression. If
$\mathbf n_m$ points from the particle centre towards neighbour $m$, the
corresponding compressive traction on the particle is $-p_m\mathbf n_m$.

The inserted-ion molar concentration is
\begin{equation}
  n=n_\ell y,
  \label{eq:SI_n_y}
\end{equation}
where $n_\ell$ is the molar concentration of lattice sites and $y\in[0,1]$ is the local
mole fraction.

Bold lower-case symbols denote vectors and bold Greek symbols denote
second-order tensors. The gradient is taken with respect to the reference
coordinate $\mathbf X$,
\begin{equation}
  (\nabla\mathbf u)_{ij}=\frac{\partial u_i}{\partial X_j},
  \qquad
  \sym\nabla\mathbf u
  =\frac12\left(\nabla\mathbf u+\nabla\mathbf u^{T}\right).
  \label{eq:SI_gradient}
\end{equation}
We use $\tr\mathbf A=A_{ii}$,
$\mathbf A:\mathbf B=A_{ij}B_{ij}$, and
$(\mathbf a\otimes\mathbf b)_{ij}=a_i b_j$. Repeated Cartesian indices are
summed. The positive-part operator is $\pos{x}=\max(x,0)$, and $\d V$,
$\d A$, and $\d\Omega$ denote volume, area, and solid-angle measures.

A particle occupies the reference ball $\mathcal B$ of radius $R$. Its
particle-average mole fraction is
\begin{equation}
  Y=\frac{1}{|\mathcal B|}\int_{\mathcal B}y\,\d V.
  \label{eq:SI_Y}
\end{equation}
The isotropic swelling strain is $g y\mathbf I$. With the thermodynamic
pressure $p=-\tr\boldsymbol\sigma/3$ and partial molar volume
$v_c=\partial\mu/\partial p$, the main-text convention gives
\begin{equation}
  g=\tfrac13 n_\ell v_c.
  \label{eq:SI_g_vc}
\end{equation}

\begin{figure}[t]
  \centering
  \includegraphics[width=0.28\textwidth]{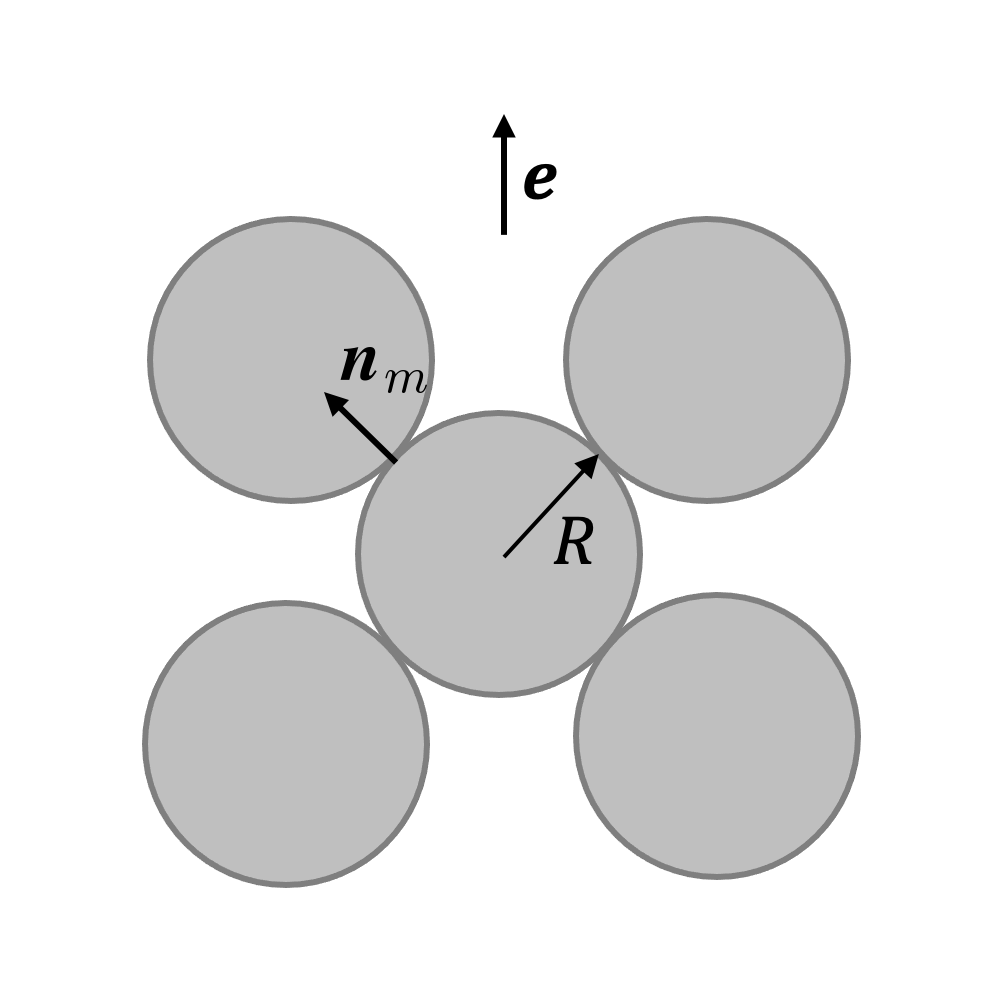}
  \caption{Planar schematic of a central particle and four neighbours from an
  FCC contact network. The particle radius is $R$, $\mathbf e$ is the
  through-cell direction, and $\mathbf n_m$ is the unit vector from the
  particle centre towards contact $m$. The complete FCC neighbourhood has
  twelve nearest neighbours; the sketch displays only the contacts lying in
  the illustrated section.}
  \label{fig:SI_geometry}
\end{figure}

Let $\{\mathbf e_1,\mathbf e_2,\mathbf e_3\}$ be a Cartesian orthonormal
basis. The idealised microstructures in the main text are periodic arrays of
identical spheres with nearest-neighbour directions
\begin{align}
  \mathcal N_{\SC}
  &=\{\pm\mathbf e_1,\pm\mathbf e_2,\pm\mathbf e_3\},
  \label{eq:SI_normals_sc}
  \\
  \mathcal N_{\BCC}
  &=\left\{\frac{(\pm1,\pm1,\pm1)}{\sqrt3}\right\},
  \label{eq:SI_normals_bcc}
  \\
  \mathcal N_{\FCC}
  &=\left\{
  \frac{(\pm1,\pm1,0)}{\sqrt2},
  \frac{(\pm1,0,\pm1)}{\sqrt2},
  \frac{(0,\pm1,\pm1)}{\sqrt2}
  \right\}.
  \label{eq:SI_normals_fcc}
\end{align}
We take the through-cell direction to coincide with one cubic axis, without
loss of generality $\mathbf e=\mathbf e_1$. The projection factor
\begin{equation}
  \alpha_m=(\mathbf n_m\cdot\mathbf e)^2
  \label{eq:SI_alpha}
\end{equation}
measures how strongly through-cell strain changes the separation of contact
pair $m$. In the touching reference geometry, the nearest-neighbour
separation is $2R$. If $a$ is the conventional cubic lattice parameter,
\begin{equation}
  a_{\SC}=2R,
  \qquad
  a_{\BCC}=\frac{4R}{\sqrt3},
  \qquad
  a_{\FCC}=2\sqrt2R.
  \label{eq:SI_lattice_parameters}
\end{equation}

The imposed lattice deformation moves particle centres according to the
small uniaxial strain $\varepsilon_e\mathbf e\otimes\mathbf e$. A reference
centre-to-centre vector $2R\mathbf n_m$ therefore becomes
\begin{equation}
  2R\left(\mathbf I+\varepsilon_e\mathbf e\otimes\mathbf e\right)
  \mathbf n_m,
  \label{eq:SI_centres_map}
\end{equation}
and its length is
\begin{equation}
  2R\left(1+\alpha_m\varepsilon_e+O(\varepsilon_e^2)\right).
  \label{eq:SI_centres_distance}
\end{equation}
Equation~\eqref{eq:SI_centres_map} updates only the prescribed lattice
geometry; the particle remains a small-strain linearly elastic solid.
Symmetric contact between two identical neighbours is represented by the
plane halfway between their deformed centres. Defining the half-separation
vector
\begin{equation}
  \mathbf d_m
  =R\left(\mathbf I+\varepsilon_e\mathbf e\otimes\mathbf e\right)
  \mathbf n_m,
\end{equation}
the plane has outward unit normal and distance from the particle centre
\begin{equation}
  \mathbf n_m^e=\frac{\mathbf d_m}{|\mathbf d_m|},
  \qquad
  d_m^e=|\mathbf d_m|.
  \label{eq:SI_contact_plane}
\end{equation}
At zero electrode strain these distances may equivalently be written as
$a/2$, $\sqrt3a/4$, and $\sqrt2a/4$ for SC, BCC, and FCC nearest-neighbour
planes, respectively; all reduce to $R$ in the touching geometry.

\section{Finite element solution of the equilibrium problem}
\label{sec:SI_FE}

\subsection{Strong form}

The numerical formulation is the stationary problem described in the main text. In the corresponding time-dependent model, inserted ions
satisfy a conservation law while the active material particle remains in mechanical
equilibrium,
\begin{equation}
  \frac{\partial n}{\partial t}+\nabla\cdot\mathbf J=0,
  \qquad
  \nabla\cdot\boldsymbol\sigma=\mathbf0,
  \label{eq:SI_time_dependent_problem}
\end{equation}
where $\mathbf J$ is the molar flux. At equilibrium the chemical potential is
spatially constant. This implies $\partial n/\partial t=0$, but not
$\nabla n=0$: a nonuniform stress can be balanced by a nonuniform filling
fraction.

The additive small-strain decomposition is
\begin{equation}
  \sym\nabla\mathbf u
  =\boldsymbol\varepsilon_{\mathrm{el}}
  +\boldsymbol\varepsilon_{\mathrm{sw}},
  \qquad
  \boldsymbol\varepsilon_{\mathrm{sw}}
  =\tfrac13 v_c n\,\mathbf I
  =g y\mathbf I.
  \label{eq:SI_strain_split}
\end{equation}
The free energy density used in the finite element formulation is
\begin{equation}
  \psi(\boldsymbol\varepsilon_{\mathrm{el}},n)
  =\psi_{\mathrm{chem}}(n)
  +\mu_L\boldsymbol\varepsilon_{\mathrm{el}}:
  \boldsymbol\varepsilon_{\mathrm{el}}
  +\frac{\lambda_L}{2}
  \left(\tr\boldsymbol\varepsilon_{\mathrm{el}}\right)^2,
  \label{eq:SI_FE_energy}
\end{equation}
where $\lambda_L$ and $\mu_L$ are the Lam\'e parameters. Consequently,
\begin{align}
  \boldsymbol\sigma
  &=2\mu_L\boldsymbol\varepsilon_{\mathrm{el}}
  +\lambda_L\tr(\boldsymbol\varepsilon_{\mathrm{el}})\mathbf I,
  \label{eq:SI_FE_hooke}
  \\
  \mu
  &=\Rgas T f(y)-\tfrac13 v_c \tr\boldsymbol\sigma,
  \qquad
  f(y)=\ln\!\left(\frac{y}{1-y}\right),
  \label{eq:SI_FE_mu}
\end{align}
where $\Rgas$ is the molar gas constant and $T$ is the absolute temperature.
Thus the chemical part used in the finite element calculations is $\mu_{\mathrm{chem}}(y)=\Rgas T f(y)$.
After scaling the chemical potential
by $\Rgas T$, this is the expression $\mu_{\mathrm{chem}}=\ln(y/(1-y))$
quoted in the main-text figure caption. The contact analysis below only
requires $\mu_{\mathrm{chem}}$ to be smooth and locally monotone at the
mole fraction considered.

For a given chemical potential $\mu$, the stationary unknowns are the
displacement $\mathbf u$ and the mole fraction $y$.
They satisfy
\begin{align}
  \nabla\cdot\boldsymbol\sigma&=\mathbf0
  &&\text{in }\mathcal B,
  \label{eq:SI_FE_equilibrium}
  \\
  \Rgas T f(y) - \tfrac13 v_c \tr \boldsymbol \sigma - \mu &= 0
  &&\text{in }\mathcal B,
  \label{eq:SI_FE_chemical_constraint}
\end{align}

\subsection{Contact and symmetry conditions}

Let $\mathbf x=\mathbf X+\mathbf u$ be the current position of a particle
point. The signed gap to contact plane $m$ is
\begin{equation}
  h_m(\mathbf X,\mathbf u)
  =d_m^e-\mathbf x\cdot\mathbf n_m^e.
  \label{eq:SI_FE_gap}
\end{equation}
The plane normal points from the particle centre towards its neighbour, so
$h_m<0$ corresponds to penetration of the rigid symmetry plane.
Frictionless unilateral contact is described by
\begin{equation}
  h_m\ge0,
  \qquad
  p_m\ge0,
  \qquad
  h_mp_m=0.
  \label{eq:SI_FE_complementarity}
\end{equation}
The finite element calculations enforce these conditions with the penalty
law
\begin{equation}
  p_m=k_{\mathrm{pen}}\pos{-h_m},
  \qquad
  \mathbf t_m=-p_m\mathbf n_m^e,
  \label{eq:SI_FE_penalty}
\end{equation}
where $k_{\mathrm{pen}}$ is the penalty stiffness and $\mathbf t_m$ is the
traction on the particle. The spherical surface is traction free outside the
active contact patches.

Lattice symmetry reduces the computational domain to the octant
\begin{equation}
  \mathcal B_+=\{\mathbf X\in\mathcal B:X_1,X_2,X_3\ge0\}.
\end{equation}
On each coordinate symmetry plane the normal displacement vanishes; the
tangential traction condition is natural. Contacts related by reflection are
reconstructed from the solution on $\mathcal B_+$.

\subsection{Weak form and nonlinear solve}

On the octant, the stationary weak problem is: find $\vec u \in V$, $y \in W$ such that
\begin{align}
  &\int_{\mathcal B_+}
  \boldsymbol\sigma:\sym\nabla\mathbf v\,\d V
  +\sum_m\int_{\Gamma_m^+}
  k_{\mathrm{pen}}\pos{-h_m}
  \mathbf n_m^e\cdot\mathbf v\,\d A=0,&&
  \quad \forall \vec{v} \in V
  \label{eq:SI_FE_weak_mechanics}
  \\
  &\int_{\mathcal B_+} \left(\Rgas T f(y)-\tfrac13 v_c\tr\boldsymbol\sigma-\mu\right) w \, \d V=0, &&
  \quad \forall w \in W
  \label{eq:SI_FE_weak_chemical}
\end{align}
where $\Gamma_m^+$ is the candidate part of the particle boundary associated
with contact plane $m$ in the octant. The vector function space $V$ is a
quadratic Lagrange space, and the scalar function space $W$ is a linear Lagrange space. The coupled nonlinear system is solved by Newton--Raphson iteration with a backtracking line search.

The full compressive load $P_m$ is obtained by integrating $p_m$ over the
contact patch and applying the appropriate reflection multiplicity when only
a fraction of the patch lies in $\mathcal B_+$. Together with the computed
value of $Y$, these loads are the numerical quantities compared
with the asymptotic theory. The contact implementation was checked directly
against the Hertz load as functions of imposed strain and mole fraction.

\section{Spherical averaging of the particle equations}
\label{sec:SI_radial_average}

The finite-element problem of Sec.~\ref{sec:SI_FE} is fully
three-dimensional because the contact tractions are localised on the particle
surface and therefore depend on angular position. For the asymptotic analysis,
however, we do not need to reconstruct this complete angular dependence.
Instead, we seek equations for spherical averages that are sufficient to
determine the contact-induced shift in the spatially uniform chemical
potential.

To make clear what must be averaged, recall that the equilibrium problem
consists of the mechanical equation
\begin{equation*}
  \nabla\cdot\boldsymbol\sigma=\mathbf0,
  \qquad\text{Eq.~\eqref{eq:SI_FE_equilibrium},}
\end{equation*}
together with the chemical-equilibrium condition
\begin{equation*}
  \mu_{\mathrm{chem}}(y)
  -\frac{v_c}{3}\tr\boldsymbol\sigma
  =\mu,
  \qquad\text{Eq.~\eqref{eq:SI_FE_chemical_constraint}.}
\end{equation*}
We therefore have two tasks. First, we must average the mechanical
equilibrium equation, which requires the radial and tangential components of
the stress. Second, we must average the chemical-equilibrium equation, which
requires the trace of the stress. We now derive these quantities in terms of
spherically averaged displacement and filling fraction.

Write
\[
  r=|\mathbf X|,
  \qquad
  \eR=\frac{\mathbf X}{r}.
\]
For any scalar $f(r,\theta,\phi)$, define its spherical average at fixed
radius by
\begin{equation}
  \sphereav{f}(r)
  =\frac{1}{4\pi}\int_{\mathbb S^2}
  f(r,\theta,\phi)\,\d\Omega,
  \qquad
  \d\Omega=\sin\theta\,\d\theta\,\d\phi.
  \label{eq:SI_spherical_average}
\end{equation}
For displacement and stress, define the averaged radial displacement,
radial stress, and mean hoop stress by
\begin{align}
  \sphereav{u}(r)
  &=\sphereav{\mathbf u\cdot\eR},
  \label{eq:SI_average_u}
  \\
  \sphereav{\sigma}_{rr}(r)
  &=\sphereav{\eR\cdot\boldsymbol\sigma\eR},
  \label{eq:SI_average_srr}
  \\
  \sphereav{\sigma}_{\theta\theta}(r)
  &=\frac12\sphereav{
  (\mathbf I-\eR\otimes\eR):\boldsymbol\sigma}.
  \label{eq:SI_average_stt}
\end{align}
The last definition is the mean of the two tangential normal stresses and
does not assume axisymmetry.

We first determine the strain quantities needed to construct these stresses.
From Eqs.~\eqref{eq:SI_strain_split} and \eqref{eq:SI_FE_hooke},
\begin{equation*}
  \boldsymbol\varepsilon_{\mathrm{el}}
  =\sym\nabla\mathbf u-gy\mathbf I,
  \qquad
  \boldsymbol\sigma
  =2\mu_L\boldsymbol\varepsilon_{\mathrm{el}}
  +\lambda_L
  \tr(\boldsymbol\varepsilon_{\mathrm{el}})\mathbf I.
\end{equation*}
The radial stress therefore requires the radial strain
$\eR\cdot(\sym\nabla\mathbf u)\eR$, while the isotropic part of the
constitutive law, and hence also $\tr\boldsymbol\sigma$ in the chemical
equilibrium equation, requires
$\tr(\sym\nabla\mathbf u)$. These are the two strain quantities that we now
average.

Decompose the displacement into radial and tangential parts,
\begin{equation*}
  \mathbf u
  =u_r\eR+\mathbf u_S,
  \qquad
  u_r=\mathbf u\cdot\eR,
  \qquad
  \mathbf u_S\cdot\eR=0.
\end{equation*}
In spherical coordinates,
\begin{equation*}
  \tr(\sym\nabla\mathbf u)
  =\nabla\cdot\mathbf u
  =
  \frac{\partial u_r}{\partial r}
  +\frac{2u_r}{r}
  +\frac{1}{r}\nabla_S\cdot\mathbf u_S,
\end{equation*}
where $\nabla_S$ denotes the surface gradient on $\mathbb S^2$. The final
term contains the angular variation of the displacement. Its spherical
average vanishes because the sphere is closed: for every regular tangential
field $\mathbf q$,
\begin{equation}
  \int_{\mathbb S^2}
  \nabla_{\!S}\cdot\mathbf q\,\d\Omega=0.
  \label{eq:SI_surface_divergence}
\end{equation}
Moreover, the radial normal strain is
\begin{equation*}
  \eR\cdot(\sym\nabla\mathbf u)\eR
  =\frac{\partial u_r}{\partial r}.
\end{equation*}
Since differentiation with respect to $r$ commutes with the angular
average, Eq.~\eqref{eq:SI_surface_divergence} therefore gives
\begin{equation}
  \sphereav{\tr(\sym\nabla\mathbf u)}
  =\sphereav{u}'+\frac{2}{r}\sphereav{u},
  \qquad
  \sphereav{\eR\cdot(\sym\nabla\mathbf u)\eR}
  =\sphereav{u}'.
  \label{eq:SI_average_strain_identities}
\end{equation}
Here and below, a prime denotes differentiation with respect to $r$.

We can now construct the averaged stress components required for mechanical
equilibrium. Substituting
$\boldsymbol\varepsilon_{\mathrm{el}}
=\sym\nabla\mathbf u-gy\mathbf I$
into the constitutive law and using
Eq.~\eqref{eq:SI_average_strain_identities} gives
\begin{align}
  \sphereav{\sigma}_{rr}
  &=(2\mu_L+\lambda_L)\sphereav{u}'
  +2\lambda_L\frac{\sphereav{u}}{r}
  -(2\mu_L+3\lambda_L)g\sphereav{y},
  \label{eq:SI_average_sigma_rr}
  \\
  \sphereav{\sigma}_{\theta\theta}
  &=\lambda_L\sphereav{u}'
  +2(\mu_L+\lambda_L)\frac{\sphereav{u}}{r}
  -(2\mu_L+3\lambda_L)g\sphereav{y}.
  \label{eq:SI_average_sigma_tt}
\end{align}
Thus the averaged radial and hoop stresses are determined entirely by
$\sphereav{u}(r)$ and $\sphereav{y}(r)$.

We now return to the first equation that we wish to reduce: mechanical
equilibrium,
Eq.~\eqref{eq:SI_FE_equilibrium}. Taking its radial projection gives
$\eR\cdot(\nabla\cdot\boldsymbol\sigma)=0$. Before averaging, this expression
contains both radial derivatives and angular derivatives of the stress.
The latter again appear as surface-divergence terms and vanish upon
integration over $\mathbb S^2$ by
Eq.~\eqref{eq:SI_surface_divergence}. The spherical average of mechanical
equilibrium is therefore
\begin{equation}
  \frac{\d\sphereav{\sigma}_{rr}}{\d r}
  +\frac{2}{r}
  \left(
    \sphereav{\sigma}_{rr}
    -\sphereav{\sigma}_{\theta\theta}
  \right)
  =0.
  \label{eq:SI_average_radial_equilibrium}
\end{equation}
This has the same form as radial force balance for a spherically symmetric
stress field, even though no spherical symmetry has been assumed for the
underlying three-dimensional solution.

Substituting Eqs.~\eqref{eq:SI_average_sigma_rr} and
\eqref{eq:SI_average_sigma_tt} into
Eq.~\eqref{eq:SI_average_radial_equilibrium} gives
\begin{equation*}
  \sphereav{\sigma}_{rr}
  -\sphereav{\sigma}_{\theta\theta}
  =
  2\mu_L
  \left(
    \sphereav{u}'-\frac{\sphereav{u}}{r}
  \right),
\end{equation*}
and hence, after differentiating
Eq.~\eqref{eq:SI_average_sigma_rr} and collecting terms,
\begin{equation}
  \sphereav{u}''
  +\frac{2}{r}\sphereav{u}'
  -\frac{2}{r^2}\sphereav{u}
  =
  \frac{2\mu_L+3\lambda_L}
       {2\mu_L+\lambda_L}
  g\sphereav{y}'.
  \label{eq:SI_average_displacement_equation}
\end{equation}
Equation~\eqref{eq:SI_average_displacement_equation} is therefore the
spherically averaged form of the mechanical equilibrium equation
\eqref{eq:SI_FE_equilibrium}.

We next reduce the second equation of interest, the chemical-equilibrium
condition \eqref{eq:SI_FE_chemical_constraint}. Taking the trace of the
constitutive law gives
\begin{equation*}
  \tr\boldsymbol\sigma
  =
  (2\mu_L+3\lambda_L)
  \left[
    \tr(\sym\nabla\mathbf u)-3gy
  \right].
\end{equation*}
Using Eq.~\eqref{eq:SI_average_strain_identities}, its spherical average is
therefore
\begin{equation*}
  \sphereav{\tr\boldsymbol\sigma}
  =
  (2\mu_L+3\lambda_L)
  \left(
    \sphereav{u}'
    +\frac{2}{r}\sphereav{u}
    -3g\sphereav{y}
  \right).
\end{equation*}
Averaging Eq.~\eqref{eq:SI_FE_chemical_constraint}, and recalling that
$\mu$ is spatially uniform at equilibrium, consequently gives
\begin{equation}
  \sphereav{\mu_{\mathrm{chem}}(y)}
  -\frac{v_c}{3}(2\mu_L+3\lambda_L)
  \left(
    \sphereav{u}'
    +\frac{2}{r}\sphereav{u}
    -3g\sphereav{y}
  \right)
  =\mu.
  \label{eq:SI_average_chemical_exact}
\end{equation}
No approximation has been made in obtaining this equation. There is,
however, one important subtlety: because the chemical contribution is
nonlinear in $y$, in general
\begin{equation*}
  \sphereav{\mu_{\mathrm{chem}}(y)}
  \ne
  \mu_{\mathrm{chem}}(\sphereav{y}).
\end{equation*}
This is the only term in the averaged equilibrium equations that is not
already expressed solely in terms of the spherical means. In
Sec.~\ref{sec:SI_small_g} we show, as part of the weak-contact expansion,
that the difference between these two quantities enters only beyond the
order required to obtain the main-text OCV correction.

We have now obtained the two averaged equilibrium equations. To complete the
radial problem, we require corresponding conditions at the particle centre
and surface, together with the prescribed total lithium content.

First, regularity at the centre requires the averaged radial displacement to
vanish:
\begin{equation}
  \sphereav{u}(0)=0.
  \label{eq:SI_average_centre_bc}
\end{equation}

Second, we convert the local contact tractions into a boundary condition for
the averaged radial stress. By definition,
\begin{equation*}
  4\pi R^2\sphereav{\sigma}_{rr}(R)
  =
  \int_{\partial\mathcal B}
  \eR\cdot\boldsymbol\sigma\eR\,\d A.
\end{equation*}
The particle surface is traction free outside the contact patches, so this
integral is determined entirely by the contacts. At contact $m$, the
compressive resultant has magnitude $P_m$ and acts approximately along the
line of centres.

To estimate the accuracy of this approximation, anticipate the weak-contact
scaling used in Sec.~\ref{sec:SI_small_g}. There the indentation satisfies
$\delta_m/R=O(g)$, while Hertz geometry gives a contact radius
$a_m\sim\sqrt{R\delta_m}$. Hence the angular radius of a contact patch is
\begin{equation*}
  \frac{a_m}{R}=O(g^{1/2}).
\end{equation*}
Across the patch, the radial direction $\eR$ therefore differs from the
contact direction by an angle $O(g^{1/2})$, so their scalar product differs
from unity by $O(g)$. Since the Hertz load itself is
$P_m=O(g^{3/2})$, the difference between the radial resultant and $P_m$ is
$O(g^{5/2})$. Summing the contributions of all contacts therefore gives
\begin{equation}
  \sphereav{\sigma}_{rr}(R)
  =
  -\frac{1}{4\pi R^2}\sum_mP_m
  +O(g^{5/2}).
  \label{eq:SI_average_surface_bc}
\end{equation}
The minus sign follows from the convention that compressive stress is
negative.

Finally, the particle-average mole fraction $Y$ defined in
Eq.~\eqref{eq:SI_Y} provides the final condition. Writing the volume integral
in spherical coordinates gives
\begin{equation*}
  Y
  =
  \frac{3}{4\pi R^3}
  \int_0^R
  \int_{\mathbb S^2}
  y(r,\theta,\phi)\,
  r^2\,\d\Omega\,\d r,
\end{equation*}
and therefore, by the definition
\eqref{eq:SI_spherical_average},
\begin{equation}
  \frac{3}{R^3}
  \int_0^R
  \sphereav{y}(r)r^2\,\d r
  =Y.
  \label{eq:SI_average_mass}
\end{equation}

We have therefore reduced the original three-dimensional equilibrium problem
to a one-dimensional problem for the spherical averages
$\sphereav{u}(r)$ and $\sphereav{y}(r)$, together with the uniform chemical
potential $\mu$. Specifically,
Eq.~\eqref{eq:SI_average_displacement_equation} is the averaged mechanical
equilibrium equation,
Eq.~\eqref{eq:SI_average_chemical_exact} is the averaged chemical
equilibrium equation, and
Eqs.~\eqref{eq:SI_average_centre_bc}--\eqref{eq:SI_average_mass} provide the
centre, surface, and lithium-content conditions.

The key benefit of this reduction is that the complicated angular structure
of each contact patch no longer needs to be resolved explicitly: to the
order required below, its effect on the averaged particle problem enters
only through the scalar contact loads $P_m$. In the next section we determine
the asymptotic size of these loads using Hertz theory and then solve the
resulting averaged problem order by order in the weak-contact limit.

\section{Weak-contact expansion and derivation of the main-text OCV law}
\label{sec:SI_small_g}

In Sec.~\ref{sec:SI_radial_average} we reduced the three-dimensional
particle problem to equations for the spherical averages
$\sphereav{u}(r)$ and $\sphereav{y}(r)$ and the spatially uniform chemical
potential $\mu$. The effect of the angularly localised contacts enters this
reduced problem through the total contact loads $P_m$, via the averaged
surface condition \eqref{eq:SI_average_surface_bc}.

Our aim in this section is to solve this averaged problem in the limit of
weak particle swelling and weak contact ($g \ll 1$). There are two steps. First, we
determine the asymptotic size of the contact loads from the geometry and
Hertz theory. Second, we expand the averaged equilibrium equations order by
order. This will show explicitly that free swelling produces no change in
chemical potential, and that the first nonzero correction arises from
contact at $O(g^{3/2})$.

\subsection{Indentation and Hertz scaling}

The dimensionless swelling coefficient $g$ defined in
Eq.~\eqref{eq:SI_g_vc} is small. We therefore take
\begin{equation}
  g\ll1,
  \qquad
  \varepsilon_e=g\widehat\varepsilon_e,
  \qquad
  \widehat\varepsilon_e=O(1).
  \label{eq:SI_small_g_scaling}
\end{equation}
The scaling $\varepsilon_e=O(g)$ is chosen because particle swelling and
the imposed electrode deformation must then both contribute to the leading
change in the separation between neighbouring particles. The material
quantities $v_c$, $\lambda_L$, $\mu_L$, and the local derivatives of
$\mu_{\mathrm{chem}}$ are held fixed as $g\to0$.

We first determine the indentation of a contacting pair. In the touching
reference configuration, the distance between neighbouring particle centres
is $2R$. Under the imposed electrode strain, Eq.~\eqref{eq:SI_centres_distance}
gives the deformed centre-to-centre distance
\begin{equation*}
  2R
  \left(
    1+\alpha_m\varepsilon_e
  \right)
  +O(\varepsilon_e^2),
\end{equation*}
where $\alpha_m$ is defined in Eq.~\eqref{eq:SI_alpha}. At leading order in
$g$, the particle undergoes free isotropic swelling by the linear strain
$gY$, so its radius becomes $R(1+gY)$. This free-swelling result will be
recovered directly from the equilibrium equations below.

The mutual indentation is the excess of the two swollen radii over the
separation of the particle centres, restricted to be non-negative.
Using the scaling \eqref{eq:SI_small_g_scaling} in
Eq.~\eqref{eq:SI_centres_distance} therefore gives
\begin{equation}
  \frac{\delta_m}{2R}
  =
  \pos{gY-\alpha_m\varepsilon_e}
  +O(g^2),
  \label{eq:SI_indentation_expansion}
\end{equation}
where $\delta_m$ is the mutual indentation of particles at contact $m$.
Thus $\delta_m/R=O(g)$ whenever that contact is closed.

We now translate this geometric indentation into a contact force. For two
identical elastic spheres, the Hertz law is
\begin{equation}
  P_m
  =
  \frac43E^*\sqrt{\frac{R}{2}}\,\delta_m^{3/2},
  \qquad
  E^*
  =
  \frac{2\mu_L(\lambda_L+\mu_L)}
       {\lambda_L+2\mu_L}.
  \label{eq:SI_Hertz}
\end{equation}
The $3/2$ exponent is important: since
Eq.~\eqref{eq:SI_indentation_expansion} gives
$\delta_m=O(g)$, Eq.~\eqref{eq:SI_Hertz} immediately implies
$P_m=O(g^{3/2})$. Contact forces are therefore asymptotically smaller than
the $O(g)$ free-swelling deformation. This confirms, self-consistently,
that contact does not modify the leading $O(g)$ swelling used to obtain
Eq.~\eqref{eq:SI_indentation_expansion}.

To make the $g$ dependence explicit, define
\begin{equation}
  Q_m
  =
  \pos{Y-\alpha_m\widehat\varepsilon_e}.
  \label{eq:SI_Qm}
\end{equation}
Equations~\eqref{eq:SI_small_g_scaling} and
\eqref{eq:SI_indentation_expansion} then give
$\delta_m=2RgQ_m+O(g^2)$, and substitution into
Eq.~\eqref{eq:SI_Hertz} yields
\begin{equation}
  P_m
  =
  g^{3/2}\widehat P_m+O(g^2),
  \qquad
  \widehat P_m
  =
  \frac83E^*R^2Q_m^{3/2}.
  \label{eq:SI_Hertz_expanded}
\end{equation}
We have therefore established the key ordering used below: the particle
swelling first enters at $O(g)$, whereas the contact load first enters at
$O(g^{3/2})$.

\subsection{Expansion of the averaged equilibrium problem}

We now return to the averaged particle problem derived in
Sec.~\ref{sec:SI_radial_average}. The quantities to be determined are
$\sphereav{u}(r)$, $\sphereav{y}(r)$, and $\mu$.

For convenience, we first restate the equations that will be expanded.
The radial stress is Eq.~\eqref{eq:SI_average_sigma_rr},
\begin{equation}
  \sphereav{\sigma}_{rr}
  =
  (2\mu_L+\lambda_L)\sphereav{u}'
  +2\lambda_L\frac{\sphereav{u}}{r}
  -(2\mu_L+3\lambda_L)g\sphereav{y}.
  \label{eq:SI_expansion_radial_stress}
\end{equation}
The averaged mechanical-equilibrium equation is
Eq.~\eqref{eq:SI_average_displacement_equation},
\begin{equation}
  \sphereav{u}''
  +\frac{2}{r}\sphereav{u}'
  -\frac{2}{r^2}\sphereav{u}
  =
  \frac{2\mu_L+3\lambda_L}
       {2\mu_L+\lambda_L}
  g\sphereav{y}'.
  \label{eq:SI_expansion_mechanical_problem}
\end{equation}
The exact averaged chemical-equilibrium equation is
Eq.~\eqref{eq:SI_average_chemical_exact},
\begin{equation}
  \sphereav{\mu_{\mathrm{chem}}(y)}
  -\frac{v_c}{3}(2\mu_L+3\lambda_L)
  \left(
    \sphereav{u}'
    +\frac{2}{r}\sphereav{u}
    -3g\sphereav{y}
  \right)
  =\mu.
  \label{eq:SI_expansion_chemical_exact}
\end{equation}
At this stage no approximation has been made to the nonlinear term
$\sphereav{\mu_{\mathrm{chem}}(y)}$.

The corresponding centre, surface, and lithium-content conditions are
Eqs.~\eqref{eq:SI_average_centre_bc},
\eqref{eq:SI_average_surface_bc}, and
\eqref{eq:SI_average_mass}:
\begin{align}
  \sphereav{u}(0)
  &=0,
  \label{eq:SI_expansion_centre_condition}
  \\
  \sphereav{\sigma}_{rr}(R)
  &=
  -\frac{1}{4\pi R^2}\sum_mP_m,
  \label{eq:SI_expansion_surface_condition}
  \\
  \frac{3}{R^3}
  \int_0^R\sphereav{y}(r)r^2\,\d r
  &=Y.
  \label{eq:SI_expansion_mass_condition}
\end{align}
In Eq.~\eqref{eq:SI_expansion_surface_condition} we have omitted the
$O(g^{5/2})$ geometric error displayed explicitly in
Eq.~\eqref{eq:SI_average_surface_bc}, since it lies beyond the
$O(g^2)$ accuracy required here.

Using the Hertz expansion \eqref{eq:SI_Hertz_expanded}, it is useful to
introduce the leading average contact pressure
\begin{equation}
  \widehat p_c
  =
  \frac{1}{4\pi R^2}
  \sum_m\widehat P_m,
  \qquad
  \frac{1}{4\pi R^2}\sum_mP_m
  =
  g^{3/2}\widehat p_c+O(g^2).
  \label{eq:SI_phat}
\end{equation}
Equation~\eqref{eq:SI_phat} makes explicit where contact enters the
perturbation problem: there is no contact contribution at
$O(1)$, $O(g^{1/2})$, or $O(g)$; the first contact traction appears at
$O(g^{3/2})$.

We next expand the unknowns. Because the Hertz load introduces the
half-integer power $g^{3/2}$, it is convenient to allow a regular expansion
in powers of $g^{1/2}$. The chemical term in
Eq.~\eqref{eq:SI_expansion_chemical_exact} is nonlinear, so it is important
to expand the \emph{pointwise} filling fraction $y$, rather than only its
spherical average:
\begin{equation}
  y(r,\theta,\phi)
  =
  y_0
  +g^{1/2}y_{1/2}
  +g y_1
  +g^{3/2}y_{3/2}
  +O(g^2).
  \label{eq:SI_pointwise_y_expansion}
\end{equation}
Taking the spherical average defined by
Eq.~\eqref{eq:SI_spherical_average} gives
\begin{equation}
  \sphereav{y}
  =
  \sphereav{y_0}
  +g^{1/2}\sphereav{y_{1/2}}
  +g\sphereav{y_1}
  +g^{3/2}\sphereav{y_{3/2}}
  +O(g^2).
  \label{eq:SI_average_y_expansion}
\end{equation}
We similarly expand the averaged displacement and the spatially uniform
chemical potential:
\begin{align}
  \sphereav{u}
  &=
  u_0
  +g^{1/2}u_{1/2}
  +g u_1
  +g^{3/2}u_{3/2}
  +O(g^2),
  \label{eq:SI_u_expansion}
  \\
  \mu
  &=
  \mu_0
  +g^{1/2}\mu_{1/2}
  +g\mu_1
  +g^{3/2}\mu_{3/2}
  +O(g^2).
  \label{eq:SI_mu_expansion}
\end{align}

To expand the first term in
Eq.~\eqref{eq:SI_expansion_chemical_exact}, we apply Taylor's theorem to
$\mu_{\mathrm{chem}}(y)$ using
Eq.~\eqref{eq:SI_pointwise_y_expansion} and then take the spherical
average. This gives
\begin{align}
  \sphereav{\mu_{\mathrm{chem}}(y)}
  &=
  \sphereav{\mu_{\mathrm{chem}}(y_0)}
  \notag\\
  &\quad
  +g^{1/2}
  \sphereav{
    \mu_{\mathrm{chem}}'(y_0)y_{1/2}
  }
  \notag\\
  &\quad
  +g
  \sphereav{
    \mu_{\mathrm{chem}}'(y_0)y_1
    +\frac12
    \mu_{\mathrm{chem}}''(y_0)y_{1/2}^2
  }
  \notag\\
  &\quad
  +g^{3/2}
  \sphereav{
    \mu_{\mathrm{chem}}'(y_0)y_{3/2}
    +\mu_{\mathrm{chem}}''(y_0)y_{1/2}y_1
    +\frac16
    \mu_{\mathrm{chem}}'''(y_0)y_{1/2}^3
  }
  +O(g^2).
  \label{eq:SI_muchem_expansion}
\end{align}
We deliberately leave this expression in its general form for the moment.
The lower-order problems will show that several of these terms vanish.

Finally, substituting Eq.~\eqref{eq:SI_average_y_expansion} into the
prescribed lithium-content condition
\eqref{eq:SI_expansion_mass_condition} gives, order by order,
\begin{equation}
  \frac{3}{R^3}
  \int_0^R\sphereav{y_0}\,r^2\,\d r
  =Y,
  \qquad
  \int_0^R\sphereav{y_\beta}\,r^2\,\d r
  =0,
  \quad
  \beta=\frac12,1,\frac32,\ldots.
  \label{eq:SI_ordered_mass}
\end{equation}
We assume throughout that
$\mu_{\mathrm{chem}}'(Y)\ne0$ and that the higher derivatives appearing in
Eq.~\eqref{eq:SI_muchem_expansion} remain bounded. The expansion therefore
applies away from a phase plateau or a stoichiometric endpoint.

We now solve Eqs.~\eqref{eq:SI_expansion_radial_stress}--%
\eqref{eq:SI_expansion_mass_condition} successively at each order. The
structure to keep in mind is simple: $O(1)$ establishes the reference
state, $O(g)$ gives stress-free swelling, and $O(g^{3/2})$ introduces the
first contact-induced stress and hence the first correction to $\mu$.

\paragraph{Analysis at $O(1)$.}

At leading order there is neither swelling, which is multiplied explicitly
by $g$ in Eqs.~\eqref{eq:SI_expansion_radial_stress} and
\eqref{eq:SI_expansion_mechanical_problem}, nor contact traction, which
starts at $O(g^{3/2})$ by Eq.~\eqref{eq:SI_phat}. Therefore
Eq.~\eqref{eq:SI_expansion_mechanical_problem} gives
\begin{equation}
  u_0''
  +\frac{2}{r}u_0'
  -\frac{2}{r^2}u_0
  =0.
  \label{eq:SI_order_zero_mechanical}
\end{equation}
The centre condition \eqref{eq:SI_expansion_centre_condition} and the
surface condition \eqref{eq:SI_expansion_surface_condition}, using the
radial stress \eqref{eq:SI_expansion_radial_stress}, become
\begin{equation}
  u_0(0)=0,
  \qquad
  (2\mu_L+\lambda_L)u_0'(R)
  +\frac{2\lambda_L}{R}u_0(R)
  =0.
  \label{eq:SI_order_zero_bc}
\end{equation}
The regular solution of Eq.~\eqref{eq:SI_order_zero_mechanical} is
$u_0=A_0r$. Substitution into
Eq.~\eqref{eq:SI_order_zero_bc} gives $A_0=0$, and hence
\begin{equation}
  u_0=0,
  \qquad
  \boldsymbol\sigma_0=\mathbf0.
  \label{eq:SI_order_zero_mechanical_result}
\end{equation}

With zero stress at this order, the original pointwise
chemical-equilibrium condition
\eqref{eq:SI_FE_chemical_constraint} reduces to
\begin{equation}
  \mu_{\mathrm{chem}}(y_0)=\mu_0.
  \label{eq:SI_order_zero_chemical_pointwise}
\end{equation}
The right-hand side is spatially uniform. Since
$\mu_{\mathrm{chem}}'(Y)\ne0$, the chemical potential is locally invertible,
so Eq.~\eqref{eq:SI_order_zero_chemical_pointwise} requires $y_0$ itself to
be spatially uniform. The leading-order mass constraint in
Eq.~\eqref{eq:SI_ordered_mass} then fixes that uniform value:
\begin{equation}
  y_0=Y,
  \qquad
  \mu_0=\mu_{\mathrm{chem}}(Y).
  \label{eq:SI_order_zero}
\end{equation}
Consequently,
\begin{equation}
  \sphereav{\mu_{\mathrm{chem}}(y_0)}
  =
  \mu_{\mathrm{chem}}(Y).
  \label{eq:SI_order_zero_muchem_average}
\end{equation}
Thus the leading-order state is simply a uniformly filled, unstressed
particle.

\paragraph{Analysis at $O(g^{1/2})$.}

We included an $O(g^{1/2})$ term in the expansions
\eqref{eq:SI_pointwise_y_expansion}--\eqref{eq:SI_mu_expansion} because the
contact problem naturally involves half-integer powers of $g$. However,
Eq.~\eqref{eq:SI_phat} shows that no contact force is present yet, and
swelling first enters at $O(g)$. The mechanical problem is therefore again
homogeneous:
\begin{equation}
  u_{1/2}''
  +\frac{2}{r}u_{1/2}'
  -\frac{2}{r^2}u_{1/2}
  =0.
  \label{eq:SI_order_half_mechanical}
\end{equation}
The centre and surface conditions obtained from
Eqs.~\eqref{eq:SI_expansion_centre_condition},
\eqref{eq:SI_expansion_surface_condition}, and
\eqref{eq:SI_expansion_radial_stress} are
\begin{equation}
  u_{1/2}(0)=0,
  \qquad
  (2\mu_L+\lambda_L)u_{1/2}'(R)
  +\frac{2\lambda_L}{R}u_{1/2}(R)
  =0.
  \label{eq:SI_order_half_bc}
\end{equation}
Exactly as at $O(1)$, Eqs.~\eqref{eq:SI_order_half_mechanical} and
\eqref{eq:SI_order_half_bc} give
\begin{equation}
  u_{1/2}=0,
  \qquad
  \boldsymbol\sigma_{1/2}=\mathbf0.
  \label{eq:SI_order_half_mechanical_result}
\end{equation}

Using $y_0=Y$ from Eq.~\eqref{eq:SI_order_zero}, the
$O(g^{1/2})$ part of the pointwise chemical-equilibrium condition
\eqref{eq:SI_FE_chemical_constraint} is
\begin{equation}
  \mu_{\mathrm{chem}}'(Y)y_{1/2}
  =
  \mu_{1/2}.
  \label{eq:SI_order_half_chemical_pointwise}
\end{equation}
As at leading order, $\mu_{1/2}$ is spatially uniform and
$\mu_{\mathrm{chem}}'(Y)\ne0$, so $y_{1/2}$ must be spatially uniform. Its
integral is zero by Eq.~\eqref{eq:SI_ordered_mass}, and therefore
\begin{equation}
  y_{1/2}=0,
  \qquad
  \mu_{1/2}=0.
  \label{eq:SI_order_half}
\end{equation}
In particular, the apparently possible quadratic contribution
$\sphereav{y_{1/2}^2}$ in
Eq.~\eqref{eq:SI_muchem_expansion} vanishes.

\paragraph{Analysis at $O(g)$.}

At $O(g)$, swelling enters the mechanical problem for the first time, but
contact does not: Eq.~\eqref{eq:SI_phat} shows that the contact traction is
still of higher order. Substituting
$y_0=Y$ from Eq.~\eqref{eq:SI_order_zero} into
Eq.~\eqref{eq:SI_expansion_mechanical_problem} gives
\begin{equation}
  u_1''
  +\frac{2}{r}u_1'
  -\frac{2}{r^2}u_1
  =
  \frac{2\mu_L+3\lambda_L}
       {2\mu_L+\lambda_L}
  \sphereav{y_0}'
  =0.
  \label{eq:SI_order_one_mechanical}
\end{equation}
At this order, the surface is still traction free. Using
Eq.~\eqref{eq:SI_expansion_radial_stress} in
Eq.~\eqref{eq:SI_expansion_surface_condition} therefore gives
\begin{equation}
  (2\mu_L+\lambda_L)u_1'(R)
  +\frac{2\lambda_L}{R}u_1(R)
  -(2\mu_L+3\lambda_L)Y
  =0.
  \label{eq:SI_order_one_bc}
\end{equation}
The regular solution of
Eq.~\eqref{eq:SI_order_one_mechanical} is $u_1=A_1r$.
Substitution into Eq.~\eqref{eq:SI_order_one_bc} gives
\begin{equation}
  A_1=Y.
  \label{eq:SI_order_one_A}
\end{equation}
Hence
\begin{equation}
  u_1(r)=Yr.
  \label{eq:SI_order_one_displacement}
\end{equation}
Using Eq.~\eqref{eq:SI_order_one_displacement},
\begin{equation}
  u_1'
  +\frac{2u_1}{r}
  -3Y
  =0,
  \qquad
  \boldsymbol\sigma_1=\mathbf0.
  \label{eq:SI_order_one_stress_free}
\end{equation}
Thus the entire $O(g)$ deformation is precisely the free isotropic swelling
associated with the eigenstrain $gY\mathbf I$ in
Eq.~\eqref{eq:SI_strain_split}. This also justifies the swollen radius
$R(1+gY)$ used in deriving
Eq.~\eqref{eq:SI_indentation_expansion}.

Because the $O(g)$ state remains stress free, and because
$y_{1/2}=0$ by Eq.~\eqref{eq:SI_order_half}, the pointwise
chemical-equilibrium condition \eqref{eq:SI_FE_chemical_constraint} reduces
at this order to
\begin{equation}
  \mu_{\mathrm{chem}}'(Y)y_1=\mu_1.
  \label{eq:SI_order_one_chemical_pointwise}
\end{equation}
Therefore $y_1$ is spatially uniform, and the $O(g)$ mass constraint in
Eq.~\eqref{eq:SI_ordered_mass} gives
\begin{equation}
  y_1=0,
  \qquad
  \mu_1=0.
  \label{eq:SI_order_one}
\end{equation}
This is an important intermediate result: \emph{free swelling by itself
does not change the equilibrium chemical potential at $O(g)$.} A
mechanically induced OCV correction can therefore appear only once contact
generates a nonzero stress.

\paragraph{Analysis at $O(g^{3/2})$.}

We now reach the first order at which contact enters. From
Eq.~\eqref{eq:SI_phat}, the averaged compressive traction on the particle
surface is $-g^{3/2}\widehat p_c+O(g^2)$.

The $O(g^{3/2})$ part of the averaged mechanical equation
\eqref{eq:SI_expansion_mechanical_problem} is
\begin{equation}
  u_{3/2}''
  +\frac{2}{r}u_{3/2}'
  -\frac{2}{r^2}u_{3/2}
  =
  \frac{2\mu_L+3\lambda_L}
       {2\mu_L+\lambda_L}
  \sphereav{y_{1/2}}'
  =0,
  \label{eq:SI_order_three_half_mechanical}
\end{equation}
where the final equality follows from
Eq.~\eqref{eq:SI_order_half}. At the surface,
Eqs.~\eqref{eq:SI_expansion_radial_stress},
\eqref{eq:SI_expansion_surface_condition}, and
\eqref{eq:SI_phat} give
\begin{equation}
  (2\mu_L+\lambda_L)u_{3/2}'(R)
  +\frac{2\lambda_L}{R}u_{3/2}(R)
  -(2\mu_L+3\lambda_L)\sphereav{y_{1/2}}(R)
  =
  -\widehat p_c.
  \label{eq:SI_order_three_half_bc}
\end{equation}
Since $\sphereav{y_{1/2}}=0$ by
Eq.~\eqref{eq:SI_order_half}, the regular solution of
Eq.~\eqref{eq:SI_order_three_half_mechanical} satisfying
Eq.~\eqref{eq:SI_order_three_half_bc} is
\begin{equation}
  u_{3/2}(r)
  =
  -\frac{\widehat p_c}
         {2\mu_L+3\lambda_L}\,r.
  \label{eq:SI_order_three_half_displacement}
\end{equation}
Consequently,
\begin{equation}
  u_{3/2}'
  +\frac{2u_{3/2}}{r}
  =
  -\frac{3\widehat p_c}
         {2\mu_L+3\lambda_L}.
  \label{eq:SI_order_three_half_divergence}
\end{equation}
Substitution of
Eq.~\eqref{eq:SI_order_three_half_displacement} into the constitutive
relations \eqref{eq:SI_expansion_radial_stress} and
\eqref{eq:SI_average_sigma_tt} shows that the spherically averaged stress
at this order is uniform and hydrostatic:
\begin{equation}
  \sphereav{\boldsymbol\sigma}_{3/2}
  =
  -\widehat p_c\mathbf I,
  \qquad
  \tr\sphereav{\boldsymbol\sigma}_{3/2}
  =
  -3\widehat p_c.
  \label{eq:SI_order_three_half_stress}
\end{equation}
This is the central mechanical result: after spherical averaging, the
localised Hertz contacts act at $O(g^{3/2})$ like a uniform hydrostatic
compression of magnitude $\widehat p_c$.

We can now determine the corresponding chemical-potential correction. The
contact traction is angularly nonuniform, so the pointwise field
$y_{3/2}$ need not be spherically symmetric. We therefore use the exact
\emph{averaged} chemical equation
\eqref{eq:SI_expansion_chemical_exact}, rather than assuming a pointwise
radial form.

Using $y_0=Y$ from Eq.~\eqref{eq:SI_order_zero},
$y_{1/2}=0$ from Eq.~\eqref{eq:SI_order_half}, and
$y_1=0$ from Eq.~\eqref{eq:SI_order_one}, the nonlinear expansion
\eqref{eq:SI_muchem_expansion} simplifies to
\begin{equation}
  \sphereav{\mu_{\mathrm{chem}}(y)}
  =
  \mu_{\mathrm{chem}}(Y)
  +g^{3/2}
  \mu_{\mathrm{chem}}'(Y)
  \sphereav{y_{3/2}}
  +O(g^2).
  \label{eq:SI_muchem_three_half}
\end{equation}
Substitution of Eq.~\eqref{eq:SI_muchem_three_half} together with the field
expansions \eqref{eq:SI_average_y_expansion},
\eqref{eq:SI_u_expansion}, and \eqref{eq:SI_mu_expansion} into the exact
averaged chemical equation \eqref{eq:SI_expansion_chemical_exact} gives at
$O(g^{3/2})$
\begin{equation}
  \mu_{\mathrm{chem}}'(Y)\sphereav{y_{3/2}}
  -\frac{v_c}{3}(2\mu_L+3\lambda_L)
  \left(
    u_{3/2}'
    +\frac{2u_{3/2}}{r}
  \right)
  =
  \mu_{3/2}.
  \label{eq:SI_order_three_half_chemical}
\end{equation}
Using the mechanical result
\eqref{eq:SI_order_three_half_divergence}, this becomes
\begin{equation}
  \mu_{\mathrm{chem}}'(Y)\sphereav{y_{3/2}}
  +v_c\widehat p_c
  =
  \mu_{3/2}.
  \label{eq:SI_order_three_half_chemical_reduced}
\end{equation}
Both $\widehat p_c$ and $\mu_{3/2}$ are independent of $r$.
Since $\mu_{\mathrm{chem}}'(Y)\ne0$,
Eq.~\eqref{eq:SI_order_three_half_chemical_reduced} therefore implies that
$\sphereav{y_{3/2}}$ is also independent of $r$. Its integral must vanish
by the $O(g^{3/2})$ mass constraint in
Eq.~\eqref{eq:SI_ordered_mass}. Hence
\begin{equation}
  \sphereav{y_{3/2}}=0,
  \qquad
  \mu_{3/2}=v_c\widehat p_c.
  \label{eq:SI_order_three_half_result}
\end{equation}
Thus the first nonzero mechanical correction to the chemical potential is
the partial molar volume multiplied by the average compressive contact
pressure.

\paragraph{Closure of the nonlinear chemical term.}

We can now justify the replacement
$\sphereav{\mu_{\mathrm{chem}}(y)}
\to\mu_{\mathrm{chem}}(\sphereav y)$ used in the main text.
Combining Eqs.~\eqref{eq:SI_average_y_expansion},
\eqref{eq:SI_order_zero},
\eqref{eq:SI_order_half}, and
\eqref{eq:SI_order_one} gives
\begin{equation}
  \sphereav y
  =
  Y
  +g^{3/2}\sphereav{y_{3/2}}
  +O(g^2).
  \label{eq:SI_average_y_reduced}
\end{equation}
Taylor expanding $\mu_{\mathrm{chem}}(\sphereav y)$ about $Y$ therefore
gives
\begin{equation}
  \mu_{\mathrm{chem}}(\sphereav y)
  =
  \mu_{\mathrm{chem}}(Y)
  +g^{3/2}
  \mu_{\mathrm{chem}}'(Y)
  \sphereav{y_{3/2}}
  +O(g^2).
  \label{eq:SI_muchem_of_average_expansion}
\end{equation}
Comparing Eq.~\eqref{eq:SI_muchem_of_average_expansion} with the expansion
of the average chemical term in
Eq.~\eqref{eq:SI_muchem_three_half} gives
\begin{equation}
  \sphereav{\mu_{\mathrm{chem}}(y)}
  =
  \mu_{\mathrm{chem}}(\sphereav y)
  +O(g^2).
  \label{eq:SI_nonlinear_average}
\end{equation}
Thus the nonlinear closure used in Eq.~(13) of the main text is not an
assumption: it follows from the weak-contact expansion and is valid through
$O(g^{3/2})$.

There is one apparent subtlety. The regular expansion above describes the
outer particle field, whereas the stress close to an individual Hertz
contact is larger and spatially localised. This does not change
Eq.~\eqref{eq:SI_nonlinear_average}. Indeed,
Eq.~\eqref{eq:SI_indentation_expansion} gives
$\delta_m/R=O(g)$, and Hertz geometry therefore gives a contact radius
$a_m/R=O(g^{1/2})$. Each contact consequently occupies an $O(g)$ fraction
of the particle surface. From Eqs.~\eqref{eq:SI_Hertz_expanded} and the
Hertz contact area $a_m^2=O(R^2g)$, the local contact pressure is
$O(g^{1/2})$. The pointwise chemical-equilibrium relation
\eqref{eq:SI_FE_chemical_constraint} then permits a local filling
perturbation of the same order, $y-Y=O(g^{1/2})$, within the contact
region. Its quadratic contribution to the spherical average of the
nonlinear chemical term is therefore
\begin{equation*}
  O(g)\times O\!\left((g^{1/2})^2\right)
  =O(g^2),
\end{equation*}
and hence lies beyond the $O(g^{3/2})$ result retained here. The local
contact region therefore does not alter
Eq.~\eqref{eq:SI_nonlinear_average} at the order of interest.

\paragraph{Resulting chemical-potential correction.}

We can now recombine the expansion. From
Eqs.~\eqref{eq:SI_u_expansion},
\eqref{eq:SI_order_zero_mechanical_result},
\eqref{eq:SI_order_half_mechanical_result},
\eqref{eq:SI_order_one_displacement}, and
\eqref{eq:SI_order_three_half_displacement},
and using Eq.~\eqref{eq:SI_phat} to return from
$\widehat p_c$ to the physical contact loads, we obtain
\begin{equation}
  \sphereav{u}(r)
  =
  \left(
    gY
    -\frac{\frac{1}{4\pi R^2}\sum_mP_m}
           {2\mu_L+3\lambda_L}
  \right)r
  +O(g^2).
  \label{eq:SI_recombined_averaged_displacement}
\end{equation}
Similarly, Eqs.~\eqref{eq:SI_mu_expansion},
\eqref{eq:SI_order_zero},
\eqref{eq:SI_order_half},
\eqref{eq:SI_order_one}, and
\eqref{eq:SI_order_three_half_result} give
\begin{align}
  \mu
  &=
  \mu_{\mathrm{chem}}(Y)
  +g^{3/2}v_c\widehat p_c
  +O(g^2)
  \notag\\
  &=
  \mu_{\mathrm{chem}}(Y)
  +\frac{v_c}{4\pi R^2}\sum_mP_m
  +O(g^2).
  \label{eq:SI_mu_contact_result}
\end{align}
Equation~\eqref{eq:SI_mu_contact_result} is the main result of the
particle-scale asymptotic calculation. Relative to the stress-free chemical
potential $\mu_{\mathrm{chem}}(Y)$, contact raises the chemical potential by
the partial molar volume multiplied by the total compressive contact load
per unit particle surface area. This is Eq.~(18) of the main text to the
stated order.

\subsection{SC, BCC, and FCC OCV corrections}

It remains to evaluate the sum over contacts in
Eq.~\eqref{eq:SI_mu_contact_result} for the three idealised lattice
geometries introduced in Eqs.~\eqref{eq:SI_normals_sc}--%
\eqref{eq:SI_normals_fcc}.

Using the Hertz law \eqref{eq:SI_Hertz} together with the indentation
\eqref{eq:SI_indentation_expansion}, the contact-induced part of
Eq.~\eqref{eq:SI_mu_contact_result} is
\begin{equation}
  \Delta\mu
  =
  \frac{2E^*v_c}{3\pi}
  \sum_m
  \pos{gY-\alpha_m\varepsilon_e}^{3/2}.
  \label{eq:SI_dmu_general}
\end{equation}
The only lattice-specific information in
Eq.~\eqref{eq:SI_dmu_general} is therefore the set of projection factors
$\alpha_m=(\mathbf n_m\cdot\mathbf e)^2$ defined in
Eq.~\eqref{eq:SI_alpha}.

For the SC lattice, Eq.~\eqref{eq:SI_normals_sc} gives four contacts
perpendicular to the through-cell direction, for which $\alpha_m=0$, and
two contacts parallel to it, for which $\alpha_m=1$. For the BCC lattice,
Eq.~\eqref{eq:SI_normals_bcc} gives eight equivalent contacts with
$\alpha_m=1/3$. For the FCC lattice,
Eq.~\eqref{eq:SI_normals_fcc} gives four contacts with $\alpha_m=0$ and
eight with $\alpha_m=1/2$.

Finally, the electrode potential and lithium chemical potential are related
by $\Delta U=-\Delta\mu/F$, where $F$ is Faraday's constant. Applying this
relation to Eq.~\eqref{eq:SI_dmu_general} gives
\begin{equation}
  \Delta U^{\mathcal L}
  =
  -\frac{2E^*v_c}{3\pi F}
  \left[
    a_{\mathcal L}(gY)^{3/2}
    +b_{\mathcal L}
    \pos{gY-c_{\mathcal L}\varepsilon_e}^{3/2}
  \right],
  \label{eq:SI_dU_lattice}
\end{equation}
where the coefficients simply count the contacts of each orientation:
\begin{equation}
\begin{array}{c|ccc}
  \mathcal L
  &a_{\mathcal L}
  &b_{\mathcal L}
  &c_{\mathcal L}
  \\ \hline
  \SC  &4&2&1\\
  \BCC &0&8&1/3\\
  \FCC &4&8&1/2
\end{array}
\label{eq:SI_lattice_coefficients}
\end{equation}
Equation~\eqref{eq:SI_dU_lattice}, together with the coefficients in
Eq.~\eqref{eq:SI_lattice_coefficients}, is Eq.~(19) of the main text.
It makes explicit how the OCV correction separates into a material factor,
through $E^*v_c$, a swelling/contact factor proportional to the
$3/2$ power of the indentation, and a purely geometric factor determined
by the orientations and number of contacts in the particle lattice.

\section{Electrode strain and through-cell stress}
\label{sec:SI_electrode_strain}

\subsection{Stress from contact forces crossing a transverse area}

The through-cell electrode stress is obtained as described in the main text: project
the contact forces onto the through-cell direction,
sum the forces crossing a representative plane normal to $\mathbf e$, and
divide by the plane area $A_\perp$. Thus
\begin{equation}
  \sigma_e^{\mathcal L}
  =-\frac{1}{A_\perp^{\mathcal L}}
  \sum_{m\in\mathcal C_{\mathcal L}}
  P_m\,|\mathbf n_m\cdot\mathbf e|,
  \label{eq:SI_projected_stress}
\end{equation}
where $\mathcal C_{\mathcal L}$ contains the contacts crossing the chosen
periodic transverse section. The minus sign follows from the convention that
compression is negative.

For a conventional cubic repeat area in the touching geometry, the required
geometric sums are
\begin{equation}
\begin{array}{c|ccc}
\mathcal L&A_\perp^{\mathcal L}
&\displaystyle\sum_{m\in\mathcal C_{\mathcal L}}
|\mathbf n_m\cdot\mathbf e|
&\alpha_m\ \text{for contributing contacts}\\ \hline
\SC&4R^2&1&1\\
\BCC&16R^2/3&4/\sqrt3&1/3\\
\FCC&8R^2&4\sqrt2&1/2
\end{array}
\label{eq:SI_transverse_geometry}
\end{equation}
The contacts with $\alpha_m=0$ contribute to the particle chemical-potential
shift but carry no through-cell force and therefore do not enter
Eq.~\eqref{eq:SI_projected_stress}.

Substitution of the Hertz loads gives
\begin{align}
  \sigma_e^{\SC}
  &=-\frac23E^*\pos{gY-\varepsilon_e}^{3/2},
  \label{eq:SI_stress_sc}
  \\
  \sigma_e^{\BCC}
  &=-\frac{2}{\sqrt3}E^*
  \pos{gY-\frac13\varepsilon_e}^{3/2},
  \label{eq:SI_stress_bcc}
  \\
  \sigma_e^{\FCC}
  &=-\frac{4\sqrt2}{3}E^*
  \pos{gY-\frac12\varepsilon_e}^{3/2}
  =-\frac23E^*\pos{2gY-\varepsilon_e}^{3/2}.
  \label{eq:SI_stress_fcc}
\end{align}
Equivalently,
\begin{equation}
  \sigma_e^{\mathcal L}
  =-\frac{2E^*}{3}A_{\mathcal L}
  \pos{B_{\mathcal L}gY-C_{\mathcal L}\varepsilon_e}^{3/2},
  \label{eq:SI_stress_compact}
\end{equation}
with
\begin{equation}
  (A_{\mathcal L},B_{\mathcal L},C_{\mathcal L})
  =
  \begin{cases}
    (1,1,1),&\mathcal L=\SC,\\
    (\sqrt3,1,1/3),&\mathcal L=\BCC,\\
    (1,2,1),&\mathcal L=\FCC.
  \end{cases}
\end{equation}
This is Eq.~(20) of the main text.

\subsection{Inversion for electrode strain and stack closure}

For either electrode $k$, the three lattice laws can be written as
\begin{equation}
  \sigma_k=-K_k\pos{g_kY_k-\alpha_k\varepsilon_k}^{3/2},
  \label{eq:SI_electrode_contact_law}
\end{equation}
where
\begin{equation}
\begin{array}{c|cc}
  \mathcal L & K_k & \alpha_k\\ \hline
  \SC  &2E_k^*/3&1\\
  \BCC &2E_k^*/\sqrt3&1/3\\
  \FCC &4\sqrt2E_k^*/3&1/2
\end{array}
\label{eq:SI_K_alpha}
\end{equation}
and the material constants may differ between electrodes. On the
closed-contact compressive branch, $\sigma_k<0$, the law can be inverted:
\begin{equation}
  \varepsilon_k(\sigma_k,Y_k)
  =\frac{1}{\alpha_k}
  \left[g_kY_k-
  \left(\frac{-\sigma_k}{K_k}\right)^{2/3}\right].
  \label{eq:SI_electrode_strain_inversion}
\end{equation}
At $\sigma_k=0$, this expression gives the contact-opening threshold
$\varepsilon_k=g_kY_k/\alpha_k$. Once contact is open, the contact-only law
has zero stress and does not determine strain; any additional tensile or
binder response lies outside this reduced model.

Let the cathode, separator, and anode have thicknesses $h_c$, $h_s$, and
$h_a$, and let $H=h_c+h_s+h_a$. If all strains are measured from the same
reference configuration, compatibility requires
\begin{equation}
  h_c\varepsilon_c+h_s\varepsilon_s+h_a\varepsilon_a
  =H\varepsilon_{\mathrm{cell}},
  \label{eq:SI_stack_compatibility}
\end{equation}
and one-dimensional equilibrium gives a common through-cell stress
\begin{equation}
  \sigma=\sigma_c=\sigma_s=\sigma_a.
  \label{eq:SI_stack_equilibrium}
\end{equation}
The separator is taken to be linear,
\begin{equation}
  \sigma_s=M_s\varepsilon_s,
  \qquad
  M_s=\lambda_s+2\mu_s,
\end{equation}
where $\lambda_s$ and $\mu_s$ are the separator Lam\'e parameters and
$M_s$ is its effective through-cell modulus under the assumed
one-dimensional kinematics. If both electrode contact networks are closed,
substitution of Eq.~\eqref{eq:SI_electrode_strain_inversion} gives
\begin{equation}
  \sum_{k\in\{a,c\}}
  \frac{h_k}{\alpha_k}
  \left[g_kY_k-
  \left(\frac{-\sigma}{K_k}\right)^{2/3}\right]
  +h_s\frac{\sigma}{M_s}
  =H\varepsilon_{\mathrm{cell}}.
  \label{eq:SI_stack_scalar}
\end{equation}
Solving this scalar equation determines the common stress and the electrode
strains. A load-controlled calculation instead prescribes $\sigma$ and uses
Eq.~\eqref{eq:SI_electrode_strain_inversion} directly. Open-contact branches
must be handled piecewise.

\section{Effect of the initial lattice spacing}
\label{sec:SI_initial_geometry}

Panels (a) and (b) of Fig.~2 in the main text use the reference geometry
adopted in the analysis: particles are just touching at $Y=0$. Contact and
the associated OCV correction therefore begin as soon as the particles
swell. Panel (c) illustrates a different initial geometry by reducing the
active-material volume fraction $\chi$ while keeping the particle radius
fixed. This dilates the lattice, introduces a finite initial gap, and delays
contact to a nonzero mole fraction.

The active material volume fractions at which identical spheres just touch
are
\begin{equation}
  \chi_{\SC}^{\mathrm{touch}}=\frac{\pi}{6},
  \qquad
  \chi_{\BCC}^{\mathrm{touch}}=\frac{\sqrt3\pi}{8},
  \qquad
  \chi_{\FCC}^{\mathrm{touch}}=\frac{\pi}{3\sqrt2}.
  \label{eq:SI_touching_volume_fractions}
\end{equation}
If a lattice is uniformly dilated to a lower volume fraction $\chi$, its
nearest-neighbour spacing is
\begin{equation}
  \ell_0^{\mathcal L}
  =2R\left(\frac{\chi_{\mathcal L}^{\mathrm{touch}}}{\chi}\right)^{1/3}.
  \label{eq:SI_spacing_from_chi}
\end{equation}
At zero imposed electrode strain, free swelling first closes the gap when
$2R(1+gY)=\ell_0^{\mathcal L}$. Hence the contact-onset filling is
\begin{equation}
  Y_{\mathrm{on}}^{\mathcal L}
  =\frac{1}{g}
  \left[
  \left(\frac{\chi_{\mathcal L}^{\mathrm{touch}}}{\chi}\right)^{1/3}-1
  \right].
  \label{eq:SI_contact_onset}
\end{equation}
Because the three touching volume fractions differ, the same chosen $\chi$
produces different onset values for SC, BCC, and FCC; if
$Y_{\mathrm{on}}^{\mathcal L}>1$, that lattice remains out of contact over
the plotted filling range. Before onset, the OCV
is the stress-free chemical curve. After onset, the same Hertzian law applies
with $gY$ replaced, to leading order, by
$g(Y-Y_{\mathrm{on}}^{\mathcal L})$. Thus changing the initial spacing shifts
the point at which the OCV curves separate but does not change the local
$3/2$ contact exponent. The larger stiffness used in panel (c) magnifies the
post-contact voltage shift; it does not determine the onset of contact.

\section{PyBaMM-based full-cell calculation}
\label{sec:SI_pybamm}

\subsection{Equilibrium electrochemical path}

We implement this coupled model in PyBaMM using the SPM. The calculation uses
PyBaMM version 26.4.2 and the \texttt{Mohtat2020} parameter set. The negative-
and positive-electrode equilibrium half-cell OCV functions and the associated
cell parameters are evaluated along a prescribed stoichiometric path, while
the mechanics is solved quasi-statically at each of 1000 cell-SOC values.

Let $z\in[0,1]$ denote full-cell SOC and
\begin{equation}
  \theta_k=\frac{c_k}{c_{k,\max}},
  \qquad k\in\{n,p\},
  \label{eq:SI_theta_definition}
\end{equation}
be the PyBaMM electrode stoichiometries. With cell area $A$, electrode
thicknesses $h_n$ and
$h_p$, active-material volume fractions $\epsilon_{s,n}$ and
$\epsilon_{s,p}$, maximum concentrations $c_{n,\max}$ and $c_{p,\max}$, and
nominal capacity $Q_{\mathrm{Ah}}$,
\begin{align}
  Q_{\mathrm{mol}}&=\frac{3600Q_{\mathrm{Ah}}}{F},
  \\
  \Delta\theta_n
  &=\frac{Q_{\mathrm{mol}}}
  {Ah_n\epsilon_{s,n}c_{n,\max}},
  \\
  \Delta\theta_p
  &=\frac{Q_{\mathrm{mol}}}
  {Ah_p\epsilon_{s,p}c_{p,\max}}.
  \label{eq:SI_pybamm_stoich_swing}
\end{align}
The initial stoichiometries are read from the initial concentrations in the
parameter set, and the path is
\begin{equation}
  \theta_n(z)=\theta_{n,0}+z\Delta\theta_n,
  \qquad
  \theta_p(z)=\theta_{p,0}-z\Delta\theta_p.
  \label{eq:SI_pybamm_stoich_path}
\end{equation}
If $U_n(\theta_n)$ and $U_p(\theta_p)$ are the two equilibrium half-cell OCV
functions supplied by \texttt{Mohtat2020}, the baseline full-cell OCV is
\begin{equation}
  U_{\mathrm{base}}(z)
  =U_p(\theta_p(z))-U_n(\theta_n(z)).
  \label{eq:SI_pybamm_baseline}
\end{equation}
The values read from or derived directly from the parameter set are
summarised in Table~\ref{tab:SI_pybamm_set}.

\begin{table}[t]
\caption{PyBaMM inputs and derived equilibrium path for the
\texttt{Mohtat2020} parameter set.}
\label{tab:SI_pybamm_set}
\begin{ruledtabular}
\begin{tabular}{lcc}
Quantity & Value & Unit\\
\hline
PyBaMM version & 26.4.2 & --\\
Parameter set & \texttt{Mohtat2020} & --\\
Cell area $A$ & $2.05\times10^{-1}$ & m$^2$\\
Negative-electrode thickness $h_n$ & $6.20\times10^{-5}$ & m\\
Separator thickness $h_s$ & $1.20\times10^{-5}$ & m\\
Positive-electrode thickness $h_p$ & $6.70\times10^{-5}$ & m\\
Total active-stack thickness $H$ & $1.41\times10^{-4}$ & m\\
Negative stoichiometry window & $0.001700\rightarrow0.838763$ & --\\
Positive stoichiometry window & $0.890701\rightarrow0.027991$ & --\\
Baseline OCV endpoints & $2.810102\rightarrow4.208644$ & V\\
Number of SOC points & 1000 & --
\end{tabular}
\end{ruledtabular}
\end{table}

The mechanics uses normalised swelling coordinates over these usable
stoichiometric windows,
\begin{equation}
  \widehat Y_n=z,
  \qquad
  \widehat Y_p=1-z.
  \label{eq:SI_pybamm_mechanical_soc}
\end{equation}
These are not the PyBaMM stoichiometries $\theta_k$; they measure the
fraction of each electrode's concentration swing relative to its
low-lithiation endpoint. The concentration swings and total free linear
swelling factors over the windows are
\begin{equation}
  \Delta c_k=c_{k,\max}|\theta_k(1)-\theta_k(0)|,
  \qquad
  g_k=\frac13\overline V_k\Delta c_k,
  \label{eq:SI_pybamm_g}
\end{equation}
where $\overline V_k$ is the chosen partial molar volume, the same physical
quantity denoted by $v_c$ in the particle-scale derivation. Thus
$g_k\widehat Y_k$ is the free isotropic linear swelling strain measured from
the low-lithiation endpoint of electrode $k$.

\subsection{Additional mechanical parameters}

The \texttt{Mohtat2020} set does not contain the particle-contact and
through-cell elastic parameters required by the present model. The additional
representative parameters used in the notebook are listed in
Table~\ref{tab:SI_pybamm_mechanics}. They are explicit modelling choices
copied from the mechanics calculation, not fitted modifications of the
\texttt{Mohtat2020} electrochemical set. The partial molar volumes determine
both the free swelling and the mechanical work per mole; the active-material
elastic constants determine the Hertz stiffness; and the separator modulus
closes the one-dimensional series-stack problem. FCC geometry is chosen as a
representative highly coordinated contact network.

\begin{table}[t]
\caption{Representative mechanical parameters not contained in \texttt{Mohtat2020}.}
\label{tab:SI_pybamm_mechanics}
\begin{ruledtabular}
\begin{tabular}{lcc}
Quantity & Value & Unit\\
\hline
Negative partial molar volume $\overline V_n$ & $8.0\times10^{-6}$ & m$^3$ mol$^{-1}$\\
Positive partial molar volume $\overline V_p$ & $1.0\times10^{-6}$ & m$^3$ mol$^{-1}$\\
Negative active-material Young modulus $E_n$ & $50$ & GPa\\
Positive active-material Young modulus $E_p$ & $150$ & GPa\\
Negative Poisson ratio $\nu_n$ & $0.30$ & --\\
Positive Poisson ratio $\nu_p$ & $0.30$ & --\\
Separator through-cell modulus $M_s$ & $1.0$ & GPa\\
Contact geometry & FCC & --
\end{tabular}
\end{ruledtabular}
\end{table}

The Lam\'e parameters and Hertz stiffness are
\begin{equation}
  \mu_k=\frac{E_k}{2(1+\nu_k)},
  \qquad
  \lambda_k=\frac{E_k\nu_k}{(1+\nu_k)(1-2\nu_k)},
\end{equation}
\begin{equation}
  E_k^*=\frac{2\mu_k(\lambda_k+\mu_k)}
  {\lambda_k+2\mu_k}.
  \label{eq:SI_pybamm_Estar}
\end{equation}
The resulting derived quantities are
\begin{equation}
\begin{array}{c|cc}
 & \text{negative} & \text{positive}\\ \hline
E_k^*\;[\mathrm{Pa}] &2.747253\times10^{10}&8.241758\times10^{10}\\
g_k &6.416594\times10^{-2}&1.017422\times10^{-2}.
\end{array}
\label{eq:SI_pybamm_derived}
\end{equation}
Particle radii are not required in this reduced calculation: $R_k$ cancels
from both the leading OCV correction and the electrode stress law.

\subsection{Clamped FCC mechanics and voltage correction}

For FCC,
\begin{equation}
  \alpha=\frac12,
  \qquad
  K_k=\frac{4\sqrt2}{3}E_k^*.
  \label{eq:SI_pybamm_FCC_constants}
\end{equation}
Let $\varepsilon_k^{\mathrm{abs}}$ denote the strain measured from the
artificial configuration in which electrode $k$ has
$\widehat Y_k=0$ and touching particles. On the closed compressive branch,
\begin{equation}
  \varepsilon_k^{\mathrm{abs}}(\sigma,\widehat Y_k)
  =\frac{1}{\alpha}
  \left[g_k\widehat Y_k-
  \left(\frac{-\sigma}{K_k}\right)^{2/3}\right].
  \label{eq:SI_pybamm_absolute_strain}
\end{equation}
At zero common stress the free through-cell strain is
$g_k\widehat Y_k/\alpha$. At $z=0$, the negative electrode has
$\widehat Y_n=0$ while the positive electrode has $\widehat Y_p=1$, so
\begin{equation}
  \varepsilon_n^{\mathrm{abs}}(0)=0,
  \qquad
  \varepsilon_p^{\mathrm{abs}}(0)=\frac{g_p}{\alpha}
  =2.034845\times10^{-2}.
  \label{eq:SI_pybamm_initial_electrode_strains}
\end{equation}
The pouch is clamped at this actual initial stress-free thickness. Therefore
``$\varepsilon_{\mathrm{cell}}=0$'' in the main text means zero change from
the $z=0$ pouch thickness, not zero strain relative to the artificial
zero-swelling contact reference.

The common stress at each $z$ is consequently determined by
\begin{align}
  &h_n\left[\varepsilon_n^{\mathrm{abs}}(\sigma,\widehat Y_n)
  -\varepsilon_n^{\mathrm{abs}}(0)\right]
  +h_s\frac{\sigma}{M_s}
  \notag\\
  &\qquad
  +h_p\left[\varepsilon_p^{\mathrm{abs}}(\sigma,\widehat Y_p)
  -\varepsilon_p^{\mathrm{abs}}(0)\right]=0.
  \label{eq:SI_pybamm_clamp}
\end{align}
Equivalently, the weighted absolute strain is held at its initial value,
corresponding to
\begin{equation}
  \varepsilon_{\mathrm{cell},0}
  =\frac{h_n\varepsilon_n^{\mathrm{abs}}(0)
  +h_p\varepsilon_p^{\mathrm{abs}}(0)}{H}
  =9.669120\times10^{-3}.
  \label{eq:SI_pybamm_initial_strain}
\end{equation}
Equation~\eqref{eq:SI_pybamm_clamp} is solved by bisection with at most 100
iterations and stopping criterion
$|\text{compatibility residual}|<10^{-10}H$. The implementation retains only
the compressive contact branch; if satisfying the closure would require
tension, the contact stress is set to zero. This safeguard is not activated
along the reported SOC path, for which the common stress is compressive for
all $z>0$. The strains plotted in the main
text are
\begin{equation}
  \varepsilon_k^{\mathrm{rel}}(z)
  =\varepsilon_k^{\mathrm{abs}}(z)
  -\varepsilon_k^{\mathrm{abs}}(0),
  \label{eq:SI_pybamm_relative_strain}
\end{equation}
with $\varepsilon_s^{\mathrm{rel}}=\sigma/M_s$ because $\sigma(0)=0$. The
absolute strains, rather than the shifted plotted strains, determine contact
indentation and therefore enter the OCV correction.

For each electrode, Eq.~\eqref{eq:SI_dU_lattice} gives
\begin{equation}
  \Delta\mu_k^{\FCC}
  =\frac{2E_k^*\overline V_k}{3\pi}
  \left[
  4(g_k\widehat Y_k)^{3/2}
  +8\pos{g_k\widehat Y_k-\frac12\varepsilon_k^{\mathrm{abs}}}^{3/2}
  \right],
  \label{eq:SI_pybamm_dmu}
\end{equation}
\begin{equation}
  \Delta U_k=-\frac{\Delta\mu_k^{\FCC}}{F},
  \qquad
  U_{\mathrm{corr}}
  =\left(U_p+\Delta U_p\right)-\left(U_n+\Delta U_n\right).
  \label{eq:SI_pybamm_voltage}
\end{equation}
As numerical checks, the notebook gives $\sigma(0)=0$, satisfies the absolute
compatibility condition to within $1.41\times10^{-14}$~m, and gives a
full-cell contact correction of $-0.744$~mV at $z=0$, rising to a maximum of
$43.443$~mV over the simulated SOC interval.

\end{document}